\documentclass[twocolumn,showpacs,preprintnumbers,prb]{revtex4}

\usepackage{amsfonts}
\usepackage{amsmath}
\usepackage{graphicx}
\usepackage{dcolumn}
\usepackage{bm}
\usepackage{float}
\usepackage{subdepth}
\usepackage{color}
\usepackage{booktabs}

\begin{document}

\newcommand{\FeO}[2]{\ensuremath{\mathrm{Fe}_{#1}\mathrm{O}_{#2}}}
\newcommand{\FeOp}[2]{\ensuremath{\mathrm{Fe}_{#1}\mathrm{O}_{#2}^{+}}}
\newcommand{\OO}{\ensuremath{\mathrm{O}_{2}}}
\newcommand{\AAA}{\r{A}}
\newcommand{\muB}{\ensuremath{\mathrm{\mu_{B}}}}
\newcommand{\cmi}{\ensuremath{\mathrm{cm}^{-1}}}

\preprint{}
\title{Geometric, electronic, and magnetic structure of $\mathbf{Fe}_{x}\mathbf{O}_y^{+}$ clusters}
\author{R.~Logemann}
\affiliation{Radboud University, Institute for Molecules and Materials, NL-6525 AJ Nijmegen, The Netherlands}
\author{G.A.~de~Wijs}
\affiliation{Radboud University, Institute for Molecules and Materials, NL-6525 AJ Nijmegen, The Netherlands}
\author{M.I.~Katsnelson}
\affiliation{Radboud University, Institute for Molecules and Materials, NL-6525 AJ Nijmegen, The Netherlands}
\author{A.~Kirilyuk}
\affiliation{Radboud University, Institute for Molecules and Materials, NL-6525 AJ Nijmegen, The Netherlands}
\date{\today }

\begin{abstract}
Correlation between geometry, electronic structure and magnetism of solids is both intriguing and elusive. This is particularly strongly manifested in small clusters, where a vast number of unusual structures appear. Here, we employ density functional theory in combination with a genetic search algorithm, GGA$+U$ and a hybrid functional to determine the structure of gas phase $\text{Fe}_{x}\text{O}_{y}^{+/0}$ clusters. 
For \FeOp{x}{y} cation clusters we also calculate the corresponding vibration spectra and compare them with experiments. We successfully identify \FeOp{3}{4}, \FeOp{4}{5}, \FeOp{4}{6}, \FeOp{5}{7} and propose structures for \FeOp{6}{8}. Within the triangular geometric structure of \FeOp{3}{4} a non-collinear, ferrimagnetic and ferromagnetic state are comparable in energy. \FeOp{4}{5} and \FeOp{4}{6} are ferrimagnetic with a residual magnetic moment of 1~\muB{} due to ionization. \FeOp{5}{7} is ferrimagnetic due to the odd number of Fe atoms. We compare the electronic structure with bulk magnetite and find \FeOp{4}{5}, \FeOp{4}{6}, \FeOp{6}{8} to be mixed valence clusters. In contrast, in \FeOp{3}{4} and \FeOp{5}{7}, all Fe are found to be trivalent. 

\end{abstract}

\maketitle
In nano technology there is an ever increasing demand for increasing the density of electronic and magnetic devices. This continuous downscaling trend drives the interest to electronic and magnetic structures at the atomic scale. In essence, two things are required: first, novel materials and building blocks with exotic physical properties. Second, a fundamental knowledge of the physical mechanism of magnetism at the sub-nanometer scale. 

Atomic clusters, having highly non-monotonous behavior as a function of size, are a promising model system to study the fundamentals of magnetism at the nanoscale and below. Such clusters consist of only tens of atoms. Quantum mechanics starts to play an essential role at this small scale, adding extra degrees of freedom. Since these clusters are studied in high vacuum, they are completely isolated from their environment. 

To use these clusters as a model system, as a starting point, a detailed understanding of the relation between their geometry and electronic structure is required. 

Even in the bulk, iron oxide has a wide variety of chemical compositions and phases with many interesting phenomena, such as the Verwey transition in magnetite. \cite{Verwey1939,Walz2002}

Experiments performed on small gas phase \FeO{x}{y} clusters beyond the two-atom case are scarce. 
The structure of one and two Fe atoms with oxygen has been studied in an argon matrix using infrared spectra. \cite{Andrews1996,Chertihin1996} The corresponding vibration frequencies have been identified using density functional theory (DFT).

Iron-oxide nanoparticles have been investigated for their potential use as catalyst in chemical reactions. \cite{Laurent2008} Furthermore, since the iron-oxygen interaction has a fundamental role in many chemical and biological processes, there have been quite some studies, both experimental and theoretical, of the chemical properties of \FeO{x}{y} clusters. \cite{Reilly2007, Wang1996, Schroder2000, Erlebach2015, Reddy2004, Reddy2004a, Fiedler1994}

The possible coexistence of two structural isomers for stoichiometric iron-oxide clusters in the size range $n\geq 5$ was experimentally measured using isomer separation by ion mobility mass spectroscopy for \FeO{n}{n} and \FeO{n}{n+1} ($n=2$-9). \cite{Ohshimo2014} Furthermore, the formation of \FeO{x}{y} clusters has been studied in the size range ($x=1$-52). \cite{Yin2009}

The number of theoretical studies is, however, manifold. The magic cluster \FeO{13}{8} was extensively studied and identified as a cluster with $C_1$ but close to $D_{4h}$ point group symmetry. \cite{Palotas2010, Sun2000, Wang1999, Kortus2000, Sun2007} However, also the geometry and electronic structure of other cluster sizes have been studied theoretically. \cite{Palotas2010, Lopez2009, Ding2009, Jones2005, Shiroishi2003, Sun2000a, Cao1998} The prediction of geometric structures requires a systematic search of the potential energy surface to find the global minimum. 

The majority of theoretical studies were performed using DFT. \cite{Erlebach2015, Erlebach2014, Reddy2004, Jones2005, Shiroishi2003, Ohshimo2014, Wang1999, Ding2009, Palotas2010, Lopez2009, Sun2000a, Sun2000, Yin2009, Reilly2007, Chertihin1996} The number of works in which \FeO{m}{n} clusters were studied with methods beyond DFT is very limited and restricted to very small cluster sizes. For \FeOp{}{} its reactivity towards $H_2$ was studied on a wave-function-based CASPT2D level. \cite{Fiedler1994} For \FeO{2}{2} the molecular and electronic structure were calculated using both DFT and wave-function-based CCSD(T) methods and a $^7B_{2u}$ ground state was found. \cite{Cao1998} Furthermore, Ref.~\onlinecite{Cao1998} reports that B3LYP functional and CCSD(T) calculations give the same energy ordering of different states, although the energy differences are overestimated by the B3LYP approach. 

Recently, the structural evolution of (\FeO{2}{3}$)_n$ nanoparticles was systematically investigated from the \FeO{2}{3} cluster towards nano particles with $n=1328$. \cite{Erlebach2014, Erlebach2015} In the size range of $n=1$-10, an interatomic potential was developed and combined with a genetic algorithm in search of the lowest-energy isomer. The isomers lowest in energy were further optimized using DFT and the hybrid functional B3LYP. This way, a systematic prediction of the cluster structure was done for neutral (\FeO{2}{3}$)_n$ clusters. 

Because of its high computational burden, in DFT the geometric structure is often only relaxed into its nearest local minimum on the potential energy surface (PES).
There is no guarantee that this local minimum corresponds to the global minimum. Almost all previous works only consider either random structures or manually constructed geometries. However, for increasing cluster size these methods become less successful in finding the lowest-energy isomer. Genetic algorithms, in which stable geometries are used to create new structures, proved to be efficient in finding the global energy minimum. \cite{Johnston2003}

This method has been successfully used for transition-metal oxide clusters. \cite{Haertelt2012,Zhai2007a}

Identification of the geometric cluster structure is a delicate and computationally demanding task. Therefore, comparison with an experimental method to confirm the theoretical findings is essential. In this work, we combine previously reported experimental vibration spectra~\cite{Kirilyuk2010} with first-principles calculations and a genetic algorithm to determine the geometric structure of cationic \FeOp{x}{y} clusters. Of the nine cluster sizes reported in Ref.~\onlinecite{Kirilyuk2010}, only the geometric structure of \FeOp{4}{6} was identified. In this work, we will also identify the geometric, electronic, and magnetic structure of \FeOp{3}{4}, \FeOp{4}{5}, \FeOp{5}{7} and propose structures for \FeOp{6}{8}. 

\section{Computational Details}
We employ a genetic algorithm (GA) as is described in Ref.~\onlinecite{Johnston2003} in combination with DFT to optimize the cluster structures. For this we use the Vienna {\it ab-initio} simulation package (\textsc{vasp}) \cite{Kresse1996} using the projector augmented wave (PAW) method. \cite{Blochl1994,Kresse1999} Since the geometry optimization is the most computationally expensive part of the genetic algorithm, we use the PBE$+U$ method\cite{Perdew1996} with limited accuracy for the genetic algorithm. For all obtained isomers low in energy, we reoptimized the geometric structure using the hybrid B3LYP functional with higher accuracy and consider different magnetic configurations. We then calculate the vibration spectra and compare them with experimental results.  

Within the DFT framework, functionals based on the local density approximation (LDA) or general gradient approximation (GGA) fail to describe strongly interacting systems such as transition-metal oxides. \cite{Anisimov2002, Anisimov1997} Due to the overestimation of the electron self-interaction, they predict metallic behavior instead of the (correct) wide-band-gap insulator. In an attempt to correct for this self-interaction, one can, for example, employ a hybrid functional, where a typical amount of $20\%$ of Hartree-Fock energy is incorporated into the exchange-correlation functional. Especially for the B3LYP functional it has been shown that this results in good agreement between the geometric structure and vibrational spectra for clusters. \cite{Haertelt2012,Burow2011,Kirilyuk2010} However, hybrid functionals are quite computationally expensive compared to LDA and GGA functionals. Therefore, in the genetic algorithm we employ the GGA$+U$ method to take into account that \FeO{}{} clusters are strongly interacting systems. We use the rotational invariant implementation introduced by Dudarev and a plane wave cutoff energy of 300~eV for these calculations. \cite{Dudarev1998} 

The differences between GGA and GGA$+U$ for iron-oxide cluster calculations have been analyzed in Ref.~\onlinecite{Palotas2010}. This study stresses the importance to go beyond GGA for transition-metal oxide clusters calculations. Aside from the well-known difference for the electronic and magnetic structure, it even finds a different lowest energy isomer than GGA for \FeO{32}{33}. In our genetic algorithm calculations we use an $U_{\text{eff}}=U-J$ of 3~eV for the Fe atoms, based on a comparison between B3LYP calculations and PBE$+U$ calculations for the smallest cluster, \FeO{3}{4} (see Sec.~\ref{sec:Fe3O4p_GGAU}).
For this comparison we also calculated the mean absolute difference ($\Delta$) between the occupied Kohn-Sham energies ($E_{i}$) using B3LYP and PBE$+U$:
\begin{equation}
\Delta = \sum_{i=1}^{n} \frac{|E_i^{\text{PBE}+U}-E_i^{\text{B3LYP}}|}{n},
\label{eq:MAD_DOS}
\end{equation}
where $n$ is the number of occupied Kohn-Sham levels. 
Note that, the binding distances are only weakly dependent on the used $U_{\text{eff}}$ and our value of 3~eV is close to values used in other works (e.g., 5~eV~\cite{Palotas2010}, 3.6~eV~\cite{Lopez2009}, 3.6~eV~\cite{Jeng2004}).

We used the genetic algorithm as described in detail in Ref.~\onlinecite{Johnston2003}. New geometries are formed by the Deaven-Ho cut and splice crossover operation. To determine the fitness we used an exponential function. A generation typically consists of 20 clusters. It has been shown that the geometry of \FeO{x}{y} clusters only weakly depends on the magnetic degree of freedom. \cite{Erlebach2014} Therefore, we restrict ourselves to the ferromagnetic case in our genetic algorithm. 

For all obtained isomers low in energy, we reoptimized the geometric structure using the hybrid B3LYP functional\cite{Becke1993}\textsuperscript{,}\footnote{In particular, we use B3LYP with the VWN3 functional as defined in Ref.~\onlinecite{Vosko1980}. } and consider all possible collinear orientations of the Fe magnetic moments by constraining the difference in majority and minority electrons. 
All forces were minimized below $10^{-3}$~eV/\AAA{}. Standard recommended PAWs with an energy cutoff of 400.0~eV are used. The clusters are placed in a periodic box of a size between 11 and 17~\AAA{}, which we checked to be sufficiently large to eliminate inter cluster interactions for each cluster size. For the cluster calculations, a single $k$-point ($\Gamma$) is used. Since we also consider cationic clusters, a positive uniform background charge is added and we correct the leading errors in the potential. \cite{Makov1995,Neugebauer1992} All simulations were performed without any symmetry constraints. The reported symmetry groups are determined afterwards within 0.03~\AAA{}. For the density of states (DOS) calculations we used a
Gaussian smearing of 0.1~eV for visual clarity.

To obtain the vibration spectra, the Hessian matrix of an optimized geometry is calculated by considering finite ionic displacements of 0.015~\AAA{} for all Cartesian coordinates of each atom. The vibration frequencies are obtained by diagonalization of the Hessian matrix. 
The absorption intensity $A_i$ is calculated using \cite{Fan1992,Porezag1996}
\begin{equation}
A_i = 974.86 g_i \left( \frac{\partial \mu}{\partial Q_i}\right),
\end{equation}
where $g_i$ is the degeneracy of the vibration mode, $Q_i$ the mass weighted vibrational mode, $\mu$ the electric dipole moment, and 974.86 an empirical factor. 
A method based on four displacements for each ion was also tested but yielded the same frequencies and absorption intensities. 
Zero-point vibrational energies (ZPVE) were calculated for the isomers lowest in energy of which the vibration spectra are also shown.


For a quantitative comparison between experimental and calculated vibrational spectra, we calculate the Pendry's reliability factor. \cite{Pendry2000} The Pendry's reliability factor is a well-established method in low-energy electron diffraction (LEED) to quantify the agreement in continuous spectra and has also been applied to vibrational spectroscopy. \cite{Rossi2010}

The experimental used infrared multiphoton dissociation method (IR-MPD) does not only depend on the absorption cross section of a vibrational mode, but also on the dissociation cross section. Therefore, we use the Pendry's reliability factor to quantify the comparison of vibration spectra since it is mainly sensitive to peak positions opposed to a comparison of squared intensity. This peak sensitivity is achieved by comparing the renormalized logarithmic derivative of the intensity $I(\omega)$:
\begin{equation}
Y(\omega) = \frac{L^{-1}(\omega)}{L^{-2}(\omega) + W^2},
\end{equation}
where $L(\omega)=I'(\omega)/I(\omega)$ and $W$ is the typical FWHM of the peaks in the spectra. The Pendry's reliability factor is defined as: 
\begin{equation}
R_{P} = \int \frac{\left[ Y_{\text{th}}(\omega)-Y_{\text{expt}}(\omega) \right]^2}{Y_{\text{th}}^2(\omega)+Y_{\text{expt}}^2(\omega)}d\omega,
\label{eq:PendryFactor}
\end{equation}
where we integrate over the experimental range of frequencies. $R_{P}$ values range from 0 to 2, where 0 means perfect agreement, 1 uncorrelated spectra, and 2 perfect anticorrelation. In practice, $R_{P}$ values of 0.3 are considered acceptable agreement within LEED. 
$Y(\omega)$ is strongly dependent on experimental noise and values close to zero, hence, we calculate $Y_{\text{expt}}(\omega)$ by fitting the experimental spectrum with multiple Lorentzian peaks and extract the corresponding $W$. The theoretical frequencies are also convoluted with Lorentzian peaks with the same $W$. $R_{P}$ is always minimized as function of a rigid shift of all theoretical frequencies. 

For the calculations on magnetite we used the \textsc{vasp} code. We used a Monkhorst grid of $6 \times 6 \times 2$ and an energy cutoff of 400~eV. We used the rotationally invariant LSDA$+U$ implementation by Lichtenstein \textit{et al. }\cite{Lichtenstein1995} with effective on-site Coulomb and exchange parameters: $U=4.5$~eV~\cite{Anisimov1996} and $J=0.89$~eV for the Fe ions.

We used the monoclinic structure as described in Refs.~\onlinecite{Jeng2004, Wright2001}, and calculated the electron density with 56 atoms in the unit cell. In Ref.~\onlinecite{Jeng2004}, the charge and magnetic moment were calculated by integrating the density and spin density in a sphere with a radius of 1~\AAA{} for Fe. This radius appears to be chosen such that comparable values with neutron and x-ray diffraction experiments were obtained. 

Note, there is no unambiguous way to define these radii in systems consisting of two or more atom types. Therefore, we checked the correspondence of our results to the earlier reported ones and also performed calculations with a larger radius of 1.3~\AAA{} for Fe and 0.82~\AAA{} for O. This is a reasonable choice for \FeOp{m}{n} clusters since the overlap between different spheres is minimal, but most of the intra cluster space is covered. 

\section{Results and discussion}

\subsection{Magnetite}\label{sec:magnetite}
Even in the bulk, iron oxide is well known for its wide variety of phases and transitions. Magnetite (\FeO{3}{4}), the most stable phase of \FeO{m}{n}, is for example well known for its Verwey transition. \cite{Verwey1939,Walz2002} Above the transition temperature $T_V$, the structure is a cubic inverse spinel. Upon cooling below $T_V$, the conductivity decreases by two orders of magnitude due to charge ordering. Furthermore, the structure changes to monoclinic. 

Magnetite has the formal chemical formula ($\text{Fe}_{A}^{3+}[ \text{Fe}^{2+},\text{Fe}^{3+} ]_{B}\text{O}_{4}$) where tetrahedral $A$ sites are occupied by $\text{Fe}^{3+}$ and $B$ sites contain both divalent ($\text{Fe}^{2+}$) and trivalent ($\text{Fe}^{3+}$) iron atoms. Since magnetite is a mixed valence system, it is an excellent reference system for our cluster calculations to determine their valence state and corresponding magnetic moment. 



\begin{table}[h]
\caption{Spin moments within atomic spheres of 1.3~\AAA{} for the Fe ions in monoclinic \FeO{3}{4}. For reference the values within a sphere of 1.0~\AAA{} are also shown. $\text{A}$ and $\text{B}$ labels are consistent with Ref.~\onlinecite{Jeng2004}. }
\label{tab:magnetite_moments}
\begin{ruledtabular}
\begin{tabular}{ldd}
Site & \multicolumn{1}{c}{Spin moment ($\mu_B$)} & \multicolumn{1}{c}{Spin moment ($\mu_B$)} \\
Radius sphere & 1.3~\text{\AA} & 1.0~\text{\AA} \\ 
\hline
$\text{Fe}^{3+}$(A) & -4.02 & -3.78\\
$\text{Fe}^{2+}$(B1) & 3.69 & 3.45\\
$\text{Fe}^{3+}$(B2) & 4.15 & 3.93\\
$\text{Fe}^{3+}$(B3) & 4.06 & 3.84\\
$\text{Fe}^{2+}$(B4) & 3.64 & 3.40\\
\bottomrule
\end{tabular}
\end{ruledtabular}
\end{table}

In Table~\ref{tab:magnetite_moments}, the spin moments are shown for the different iron ions. The magnetic moments on the $A$ and $B$ sites are antiparallel creating a ferrimagnetic structure. Within the atomic spheres of 1.3~\AAA{} the $\text{Fe}^{2+}$ and $\text{Fe}^{3+}$ ions have a distinct magnetic moment of 4.0~\muB{} and 3.7~\muB{} respectively. Note the difference of 0.3~\muB{} is much smaller than the 1~\muB{} atomic value and does not depend on the size of the atomic sphere used in the range between 1.0 and 1.3~\AAA{}.

\subsection{GGA+U}\label{sec:Fe3O4p_GGAU}
\begin{figure}[ht]
\includegraphics[width=8cm]{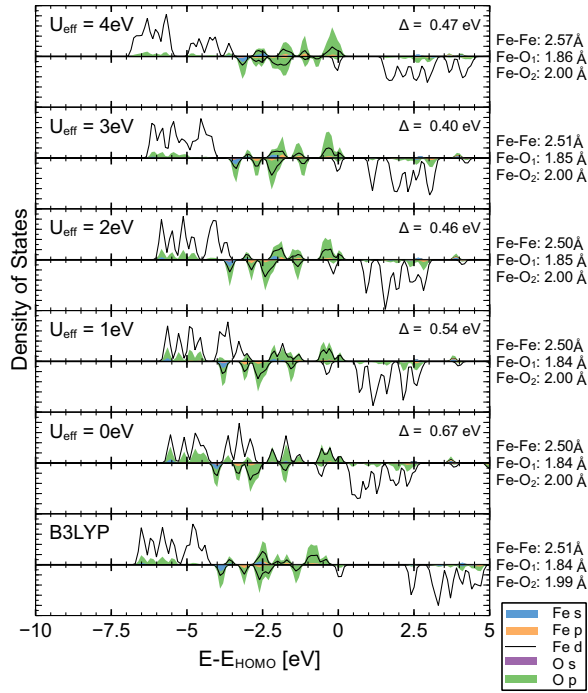}
\caption{(Color online) The density of states for the hybrid B3LYP functional and PBE$+U$ for different values of $U_{\text{eff}}$. The average inter atomic distances are shown on the right, where Fe-O$_1$ and Fe-O$_2$ refer to the Fe-O distances between bridging O atoms (side) and the capping O atom (center), respectively. The mean absolute difference $\Delta$ [Eq.~\ref{eq:MAD_DOS}] between the PBE$+U$ and B3LYP energy levels is also shown and is minimal for $U_{\text{eff}} = 3$~eV, indicating the best match in DOS.
} 
\label{fig:dos_pbeu}
\end{figure}

To determine the optimal $U_{\text{eff}}$ in comparison to the B3LYP functional for the genetic algorithm, we performed PBE$+U$ calculations on the neutral \FeO{3}{4} cluster. The results for the electronic DOS are shown in Fig.~\ref{fig:dos_pbeu} and compared with the hybrid B3LYP functional. 

The valence states within -4 and 0~eV are formed by hybridized orbitals between the $d$ orbitals of iron and the $p$ orbitals of oxygen. For increasing $U$, the majority spin $d$ orbitals of Fe decrease in energy, whereas HOMO-LUMO gap increases. Note that the HOMO-LUMO gap of 1.5~eV for $U_{\text{eff}} = 4$~eV still is 0.9~eV smaller than the 2.4~eV gap for B3LYP. Furthermore, for $U_{\text{eff}} = 2$ and 3~eV the Fe $d$ DOS features are very similar to those of the B3LYP result. To quantify this we also calculated the mean absolute difference $\Delta$ [Eq.~\ref{eq:MAD_DOS}] for the occupied levels; the results are shown in Fig.~\ref{fig:dos_pbeu}. $\Delta$ is minimal for $U_{\text{eff}} = 3$~eV, indicating the best DOS correspondence to B3LYP. We also show the corresponding bonding distances within the cluster, where Fe-O$_1$ and Fe-O$_2$ refer to the Fe-O distances between bridging O atoms (side) and the capping O atom (center), respectively. Note the interatomic distances only change very little with increasing $U_{\text{eff}}$. For $U_{\text{eff}}=3$~eV, the binding distances are within 0.01~\AAA{}; furthermore, for $U_{\text{eff}} = 3$~eV and B3LYP the occupied $d$ orbitals of Fe are at comparable energies with respect to the HOMO level. 
We therefore used $U_{\text{eff}}=3$~eV for our genetic algorithm calculations. 

\subsection{$\mathbf{Fe}_{3}\mathbf{O}_{4}^{0}$}
\begin{figure}[ht]
\includegraphics[width=8cm]{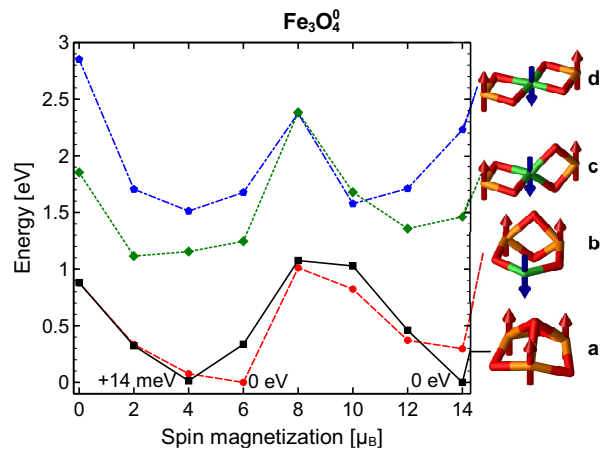}
\caption{(Color online) The energy as function of spin magnetization for different neutral \FeO{3}{4} isomers. The geometric figures on the right show the corresponding geometric structure. O atoms are shown in red, Fe spin up and Fe spin down are indicated with orange (red) and green (blue) colors (arrows), respectively. For the lowest magnetic states the relative energy differences are also shown in black. Isomers {\bf (a)} (black line) and {\bf (b)} (red line) are equally low in energy with a ferrimagnetic and ferromagnetic ground state, respectively (0~eV). The $M = 6$~\muB{} state of isomer {\bf (a)} is 14~meV higher in energy. 
} 
\label{fig:Energy_Fe3O4n}
\end{figure}

Although the possible number of isomers increases rapidly with cluster size, for small systems such as \FeO{3}{4} the number of possibilities is still small. In \FeO{3}{4}, the Fe atoms can either form a triangle or a chain. For the triangular configuration, two isomers are low in energy. 
The first isomer consists of a ring like structure where the O atoms occupy bridging states and one O atom caps the Fe triangle as is shown in Fig.~\ref{fig:Energy_Fe3O4n}{\bf (a)}. 
In the second isomer, the additional O atom is not located above the center but forms an extra bridge between the two ferromagnetic (FM) ordered Fe atoms as is shown in Fig.~\ref{fig:Energy_Fe3O4n}{\bf (b)}.

Figure~\ref{fig:Energy_Fe3O4n} shows the energy as a function of spin magnetic moment for the neutral \FeO{3}{4} cluster with four different isomers. For all spin magnetizations, the geometric structure is optimized and shown on the right with its magnetic structure lowest in energy. In Fig.~\ref{fig:Energy_Fe3O4n} and the rest of this work, Fe spin up and Fe spin down are indicated with orange (red) and green (blue) colors (arrows), respectively. O atoms are shown in red. For the neutral cluster, the two triangular isomers are equally low in energy with two different magnetic configurations. The difference is smaller than 1~meV and therefore beyond the accuracy of DFT. 
In isomer {\bf(a)}, as indicated by the black line in Fig.~\ref{fig:Energy_Fe3O4n}, the magnetic ground state corresponds to ferromagnetic alignment between the magnetic moments on the Fe atoms and a total magnetic moment of 14~\muB{}. The Fe-Fe distances are 2.51~\AAA{}, the Fe-O distances for the bridging O atoms and capping O atom are 1.84 and 1.99~\AAA{}, respectively. Aside from the FM ground state, also the ferrimagnetic state with a spin magnetization of 4~\muB{} is low in energy and only 14~meV higher than the ferromagnetic state. Note we also considered a noncollinear magnetic state with $M=0$~\muB{}, but this magnetic configuration did not turn out to be energetically stable. 

Isomer {\bf (b)} is equally low in energy and shown in red in Fig.~\ref{fig:Energy_Fe3O4n}. The magnetic ground state corresponds to a ferrimagnetic alignment where the two ferromagnetically aligned Fe atoms have Fe-O-Fe angles of approximately $90^{\circ}$. 

We also considered zero point vibrational energies for the three lowest-energy levels. When we include these into our consideration, the ferromagnetic state, indicated by the black line, is lowest in energy, and the $M=4$~\muB{} and $M=6$~\muB{} states are 17 and 19~meV higher in energy, respectively. 

\subsection{$\mathbf{Fe}_{3}\mathbf{O}_{4}^{+}$}
\begin{figure}[ht]
\includegraphics[width=8cm]{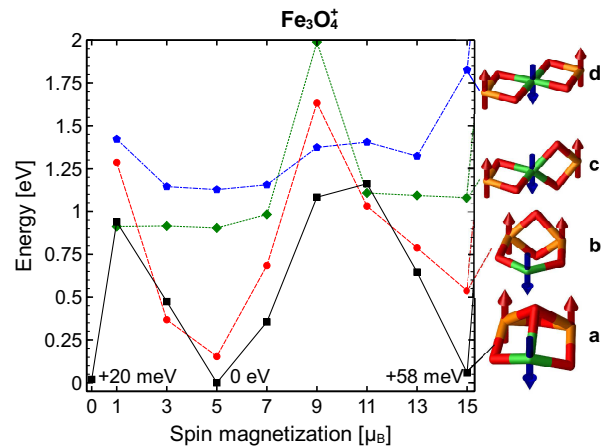}
\caption{(Color online) Energy of the \FeOp{3}{4} isomers as function of spin magnetization. Figures on the right indicate the corresponding structure. The isomer lowest in energy {\bf (a)} is a Fe triangle with three bridge O atoms and one O atom capping the triangle. For this isomer, the ferrimagnetic 5~\muB{} state is lowest in energy. The antiferromagnetic 0~\muB{} and ferromagnetic 15~\muB{} state are 20 and 58~meV higher in energy, respectively. 
Note the antiferromagnetic 0~\muB{} state corresponds to a non-collinear orientation with 120$^{\circ}$ angles between the spins. }
\label{fig:energy_Fe3O4p}
\end{figure}

\begin{figure}[ht]
\includegraphics[width=8cm]{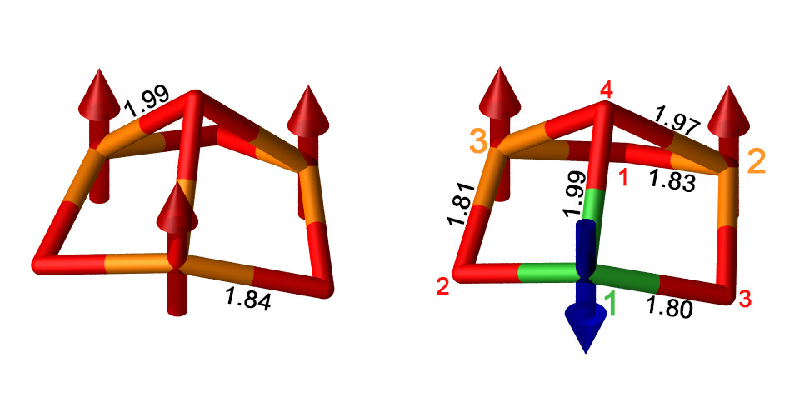}
\caption{(Color online) The neutral (left) and cation (right) \FeO{3}{4}  lowest-energy isomers. Fe spin up and Fe spin down are indicated with orange (red) and green (blue) colors (arrows), respectively. O atoms are shown in red. The interatomic distances are shown in black. The neutral and cation cluster have $C_{3v}$ and $C_{v}$ point group symmetry, respectively. } 
\label{fig:Fe3O4np_struc}
\end{figure}

For the cation \FeOp{3}{4} cluster we also considered ring and chain configurations with different oxygen locations. For all four isomers we calculated all possible different collinear magnetic states. 
Since an antiferromagnetic (AFM) triangle is the most simple example of geometrically frustrated magnetism, we also considered the non-collinear state with $M=0$~\muB{} where all magnetic moments have $120^{\circ}$ angles with respect to each other. The results are shown in Fig.~\ref{fig:energy_Fe3O4p}. 
For the charged \FeOp{3}{4} cluster, the isomer with a Fe triangle where the fourth O atom caps the triangle is, like in the neutral cluster, lowest in energy, as is shown in Fig.~\ref{fig:Fe3O4np_struc}. Three magnetic states are low in energy: 0, 5 and 15~\muB{}, with the $M=5$~\muB{} state being lowest in energy, and the non-collinear 0~\muB{} and ferromagnetic 15~\muB{} are 20~meV and 58~meV higher in energy respectively. 

The ferrimagnetic state which is lowest in energy, has a reduced symmetry ($C_{v}$) with respect to the ferromagnetic state ($C_{3v}$) and the antiferromagnetic state. This could indicate a Jahn-Teller distortion, but could also be the result of the inability of DFT to correctly model the antiferromagnetic ground state. \cite{Cramer2009,Reiher2007} However, to distinguish between these two cases, methods beyond DFT such as CASPT2 and CCSD(T) are required and therefore beyond the scope of this work. Note that different magnetic states only lead to minor differences in the vibrational frequencies.

Interestingly, the typical classical displacement during a zero-point vibration in these clusters is of the order of 0.03~\AAA{}. This is of the same order as the typical difference in inter atomic distances between different magnetic states. Therefore, this could lead to interesting phenomena in which, for example, there is a strong coupling through exchange between vibrations and magnetism. 

\begin{figure}[ht]
\includegraphics[width=8cm]{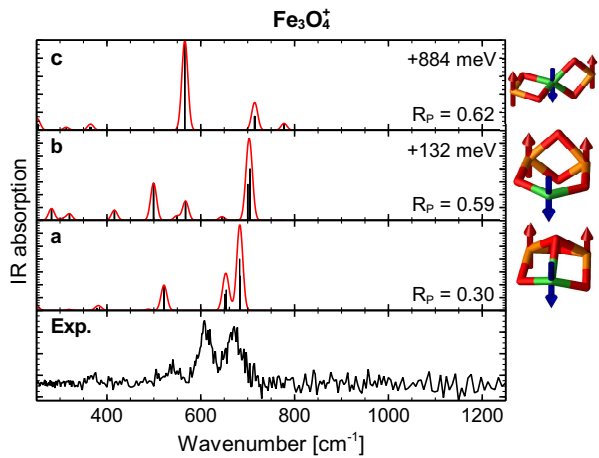}
\caption{(Color online) The experimental vibration spectra of \FeOp{3}{4} and the calculated isomers lowest in energy. The reported energy differences include ZPVE. The Pendry's reliability factor [Eq.~\ref{eq:PendryFactor}] is also shown for each isomer. 
} 
\label{fig:spectra_Fe3O4}
\end{figure}

The second triangular isomer of \FeOp{3}{4} is 154~meV higher in energy and also consists of a ring structure. The magnetic state lowest in energy has a magnetic moment of 5~\muB{}. The Fe-Fe bonding distances are 2.5 and 3.0~\AAA{} between the AFM and FM bonds within the structure. The Fe-O distances vary between 1.7 and 1.9~\AAA{}. The isomer has a $C_{2v}$ point group symmetry. 

The third and fourth isomers consist of a linear chain of Fe atoms with two O bridging atoms between each Fe pair. The two planes can be parallel or perpendicular, where the latter is lower in energy. Both isomers have a magnetic moment of 5~\muB{}. 

In Fig.~\ref{fig:spectra_Fe3O4}, both the experimental and calculated vibration spectra for the different isomers are shown. The experimental spectrum consists of three peaks at 540, 610 and 670~\cmi{}. The best match is given by isomer {\bf (a)} with calculated vibrations at 505, 630 and 660~\cmi{} and a corresponding lowest-$R_{P}$ factor of $0.30$, indicating a reasonable match with the experimental spectrum. Since isomer {\bf (a)} is also the lowest in energy, it is identified as the experimentally observed structure.  

\subsection{$\mathbf{Fe}_{4}\mathbf{O}_{5}^{0/+}$}
\begin{figure}[ht]
\includegraphics[width=8cm]{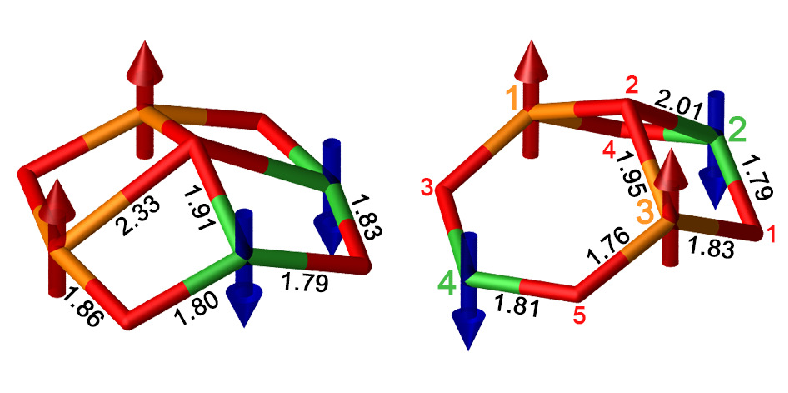}
\caption{(color online) The neutral (left) and cation (right) \FeO{4}{5}  lowest energy isomers. 
The neutral cluster has $C_{2}$ symmetry, whereas the cation cluster has $C_{s}$ symmetry. } 
\label{fig:Fe4O5np_struc}
\end{figure}

\FeO{4}{5} also consists of a ring structure in which the O atoms occupy the bridging sites and one O atom is located above the center, as is shown in Fig.~\ref{fig:Fe4O5np_struc}. The cluster has antiferromagnetic order. However, not all Fe-Fe bonds are antiferromagnetic, but also two ferromagnetically aligned bonds are present. Therefore, the cluster has no $C_{2v}$ point group symmetry but $C_{2}$, since Fe-Fe and Fe-O distances vary between 2.72-2.74~\AAA{} and 1.79-2.33~\AAA{} respectively. The magnetic state with four AFM Fe-Fe bonds is 308~meV higher in energy. 

For \FeOp{4}{5} the isomer lowest in energy consists of the same ring structure but is more symmetry broken, since the O atom above the ring is off-center as is shown in Fig.~\ref{fig:Fe4O5np_struc}. Therefore the two Fe-Fe distances are 2.69 and 3.07~\AAA{}, the Fe-O distances vary between 1.76 and 2.01~\AAA{}. The isomer has $C_{s}$ point group symmetry. Two $\mathrm{Fe}_{2}\mathrm{O}_{2}$ squares are present within the cluster. Isomer~{\bf (a)} has a magnetic moment of 1~\muB{} due to ionization. Interestingly, the ionized cluster has a different magnetic ground state with four AFM Fe-Fe bonds opposed to the neutral cluster. 

\begin{figure}[ht]
\includegraphics[width=8cm]{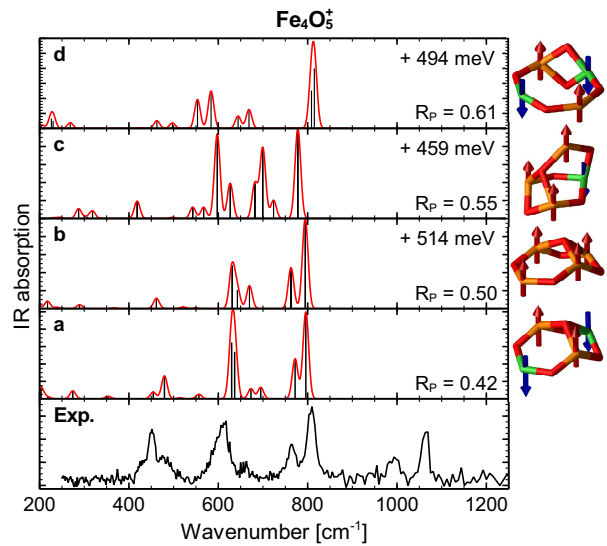}
\caption{(Color online)  The experimental and calculated vibration spectra of \FeOp{4}{5}. The isomer shown in {\bf (a)} is both the lowest in energy and  $R_{P}$~[Eq.~\ref{eq:PendryFactor}] and can therefore be identified as the experimentally observed geometrical structure. The reported energy differences include ZPVE. } 
\label{fig:spectra_Fe4O5}
\end{figure}

In Fig.~\ref{fig:spectra_Fe4O5}{\bf (b)}, we also show the vibration spectrum of the ferromagnetic state of this cluster. The Fe-Fe distances are increased to 2.74 and 3.11~\AAA{}, respectively. The ferromagnetic structure is 514~meV higher in energy. The vibration spectrum is similar but slightly shifted to the blue due to the increased bonding distances.

The second isomer, 459~meV higher in energy, is shown in Fig.~\ref{fig:spectra_Fe4O5}{\bf (c)}. This cage-like structure has $C_v$ point group symmetry and a magnetic moment of 9~\muB{}. 
Figure~\ref{fig:spectra_Fe4O5}{\bf (d)} shows the third isomer which is 494~meV higher in energy compared to Fig.~\ref{fig:spectra_Fe4O5}{\bf (a)}. The isomer has almost no symmetry ($C_{1}$), and consists of a ring where one Fe-Fe bond has two bridging O atoms. The Fe-Fe binding distances vary between 2.62 and 3.13~\AAA{}. The isomer has a magnetic moment of 1~\muB{}. 

In the experimental vibration spectrum of \FeOp{4}{5} shown in Fig.~\ref{fig:spectra_Fe4O5}, five vibration frequencies can be observed: 450, 615, 760, 810, and 1070~\cmi{}. The vibration at 1070~\cmi{} can be identified as a shifted vibration in the \OO{} messenger attached to the cluster-messenger complex and is therefore omitted in the $R_{P}$ calculation. \cite{Andrews1996} The best fit is given by isomer Fig.~\ref{fig:spectra_Fe4O5}{\bf (a)} with $R_{P} = 0.42$, which is also the isomer lowest in energy. The calculated frequencies: 479, 630, 637, 772 and 796~\cmi{} match all within 30~\cmi{} to the experimental spectrum. Also, the relative intensities between different vibrations are very similar.  Although the ferromagnetic order increases the binding distances within the cluster, the changes in the vibration spectrum of Fig.~\ref{fig:spectra_Fe4O5}{\bf (b)} are small and therefore the structure corresponding to Figs.~\ref{fig:spectra_Fe4O5}{\bf (a)} and \ref{fig:spectra_Fe4O5}{\bf (b)} can be identified as the experimentally observed structure and the IR-MPD method is not able to resolve the magnetic state in this case. 

\subsection{$\mathbf{Fe}_{4}\mathbf{O}_{6}^{0/+}$}
\begin{figure}[ht]
\includegraphics[width=8cm]{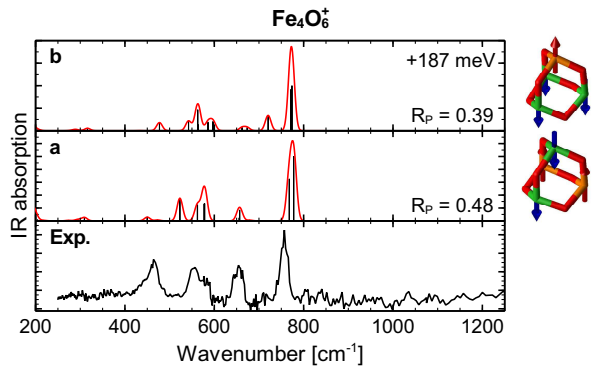}
\caption{(color online) The experimental and calculated vibration spectra of \FeOp{4}{6} for both the previous and new magnetic ground state. The vibration frequencies are very similar but differ in absorption intensity. The $M=1$~\muB{} state in {\bf (a)} is 187~meV lower in energy. } 
\label{fig:spectra_Fe4O6}
\end{figure}

In Ref.~\cite{Kirilyuk2010}, the \FeOp{4}{6} cluster was already identified as the structure shown in Fig.~\ref{fig:spectra_Fe4O6}{\bf (b)}. The reported magnetic structure was ferrimagnetic with a magnetic moment of 9~\muB{}. 

\begin{figure}[ht]
\includegraphics[width=8cm]{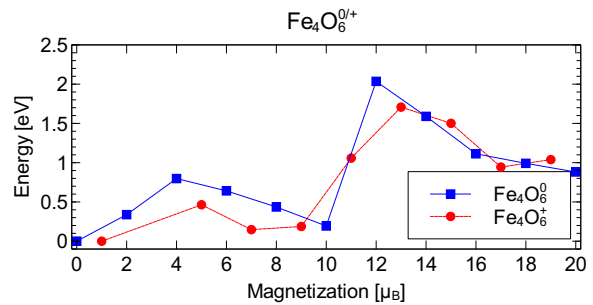}
\caption{(color online) Energy as function of magnetization of the neutral \FeO{4}{6} and cationic \FeOp{4}{6} clusters. The magnetic ground state corresponds to a total spin magnetic moment of $M=0$ and $M=1$~\muB{} for \FeO{4}{6}, and \FeOp{4}{6} respectively. }
\label{fig:energy_Fe4O6}
\end{figure}

In our calculations a magnetic state lower in energy was found for the same geometric structure for both \FeO{4}{6} and \FeOp{4}{6}. In this state \FeO{4}{6} and \FeOp{4}{6} have a magnetic moment of 0 and 1~\muB{} respectively as is shown in Fig.~\ref{fig:energy_Fe4O6}. These structures are 194 and 187~meV lower in energy for \FeO{4}{6} and \FeOp{4}{6} in comparison to the previously reported state. \cite{Kirilyuk2010} The antiferromagnetic magnetic ground state of \FeO{4}{6} was also previously reported in Ref.~\onlinecite{Erlebach2014}. For \FeO{4}{6} we also calculated a noncollinear state where all magnetic moments point towards the center of mass, such state with $M=0$~\muB{} is 30~meV higher in energy compared to the collinear $M = 0$~\muB{} state.

\begin{figure}[ht]
\includegraphics[width=8cm]{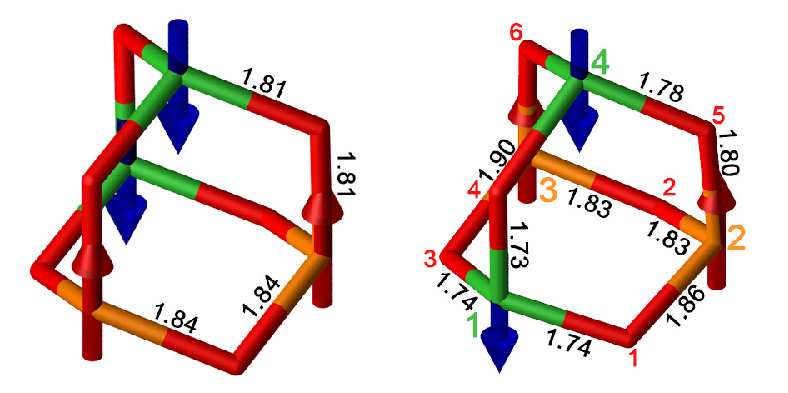}
\caption{(Color online) The neutral (left) and cation (right) \FeO{4}{6}  lowest energy isomers. 
The neutral cluster has $D_{2d}$ symmetry, whereas the cation cluster has $C_{s}$ symmetry. } 
\label{fig:Fe4O6np_struc}
\end{figure}

For the neutral cluster, minima in energy are obtained for $M=0$, 10, 20~\muB{} corresponding to flips of atomic magnetic moments of 5~\muB{} for each Fe atom. Note this also matches with an ionic picture in which the Fe atoms in \FeO{4}{6} have a $\text{Fe}^{3+}$ valence state resulting in an atomic magnetic moment of 5~\muB{}. The corresponding structure is shown in Fig.~\ref{fig:Fe4O6np_struc}. In Ref.~\onlinecite{Kirilyuk2010} is mentioned that the symmetry in the $M=10$~\muB{} state is reduced from $T_{d}$ for the ferromagnetic state to $C_{3v}$. In this antiferromagnetic ground state, the neutral cluster has $D_{2d}$ symmetry. In \FeOp{4}{6} the symmetry is reduced even further to $C_{s}$ as is shown in Fig.~\ref{fig:Fe4O6np_struc}. 

Figure~\ref{fig:spectra_Fe4O6} shows both calculated and experimental spectra for \FeOp{4}{6}. The vibration spectra for the two calculated magnetic states in Figs.~{\bf (a)} and \ref{fig:spectra_Fe4O6}{\bf (b)} show very similar behavior. The $R_{P}$ values of isomer Fig.~\ref{fig:spectra_Fe4O6}{\bf (a)} (0.48) and Fig.~\ref{fig:spectra_Fe4O6}{\bf (b)} (0.39) are both large and indicate a better match for isomer Fig.~\ref{fig:spectra_Fe4O6}{\bf (b)}. Although the spectra for Figs.~\ref{fig:spectra_Fe4O6}{\bf (a)} and \ref{fig:spectra_Fe4O6}{\bf (b)} are very similar, the ferrimagnetic structure has an extra vibration at 720~\cmi{} with small IR absorption. Furthermore, around 550~\cmi{}, vibrations differ slightly in frequency. Since the mentioned differences cannot be experimentally resolved, the IR-MPD method is unable to resolve between different magnetic states and another type of experiments such as Stern-Gerlach deflection is required to determine the magnetic moment. 

\subsection{$\mathbf{Fe}_{5}\mathbf{O}_{7}^{0/+}$}
\begin{figure}[ht]
\includegraphics[width=8cm]{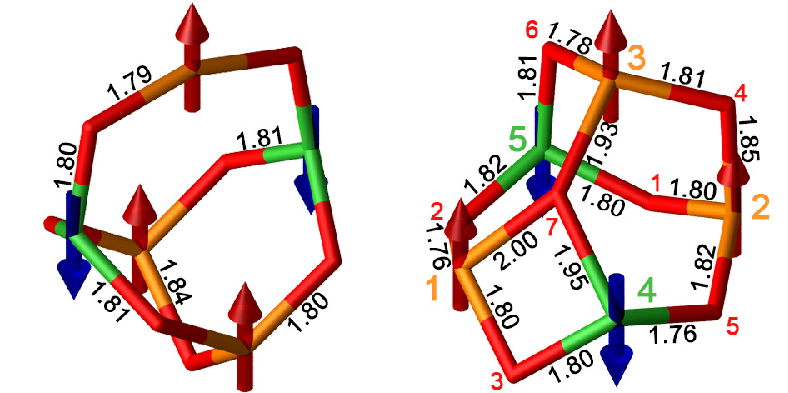}
\caption{(Color online) The neutral (left) and cation (right) \FeO{5}{7}  lowest-energy isomers. 
The neutral cluster has $C_{2v}$ symmetry, whereas the cation cluster has no symmetry. } 
\label{fig:Fe5O7np_struc}
\end{figure}

The neutral \FeO{5}{7} cluster has a ``basket" geometry as is shown in Fig.~\ref{fig:Fe5O7np_struc}. The magnetic ground state is ferrimagnetic with a total moment of 4~\muB{} due to the odd number of Fe atoms. The cluster has $C_{2v}$ symmetry. 

The cationic structure of \FeOp{5}{7} is very different and shown in Fig.~\ref{fig:Fe5O7np_struc}. Like \FeOp{4}{6}, it consists of a cage-like structure. The Fe-Fe distances range from 2.7 to 3.1~\AAA{}. Except for the triple bound O atom, all O atoms form bridges between two Fe atoms. The ground state has a magnetic moment of 5~\muB{}. 
\begin{figure}[ht]
\includegraphics[width=8cm]{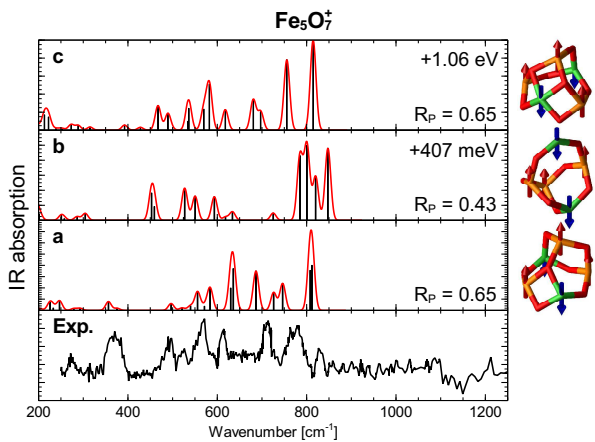}
\caption{(Color online) The experimental and calculated vibration spectra of \FeOp{5}{7}. The reported energy differences include ZPVE. } 
\label{fig:spectra_Fe5O7}
\end{figure}
The second isomer is similar to the neutral "basket" structure and is 394~meV higher in energy as is shown in Fig.~\ref{fig:spectra_Fe5O7}{\bf (b)}. The structure has $C_{s}$ symmetry and a magnetic moment of 5~\muB{}. However, the atomic spin moments have a different arrangement for the neutral and cationic state. 

The third isomer is shown in Fig.~\ref{fig:spectra_Fe5O7}{\bf (c)} and is 1.04~eV higher in energy. It contains two triple bonded O atoms and is ferrimagnetic with $M = 5$~\muB{}. 

The experimental vibration spectrum shown in Fig.~\ref{fig:spectra_Fe5O7} has eight distinct vibrations at 375, 490, 520, 570, 615, 710, 780, and 830~\cmi{} which are best resembled by the isomer lowest in energy shown in Fig.~\ref{fig:spectra_Fe5O7}{\bf (a)}, although the gap between 615 and 710~\cmi{} seems to be underestimated. Note that this also explains the high-$R_{P}$ factor of 0.65 for isomer Fig.~\ref{fig:spectra_Fe5O7}{\bf (a)}. Similar to \FeOp{4}{5} and \FeOp{4}{6} the absorption intensities of vibrations in the range of 300-500~\cmi{} are systematically underestimated.  The individual vibrations of isomer Fig.~\ref{fig:spectra_Fe5O7}{\bf (a)} are all in agreement within 35~\cmi{}. Although isomer Fig.~\ref{fig:spectra_Fe5O7}{\bf (b)} has a lower $R_{P}$ = 0.43, the energy difference of 407~meV with isomer Fig.~\ref{fig:spectra_Fe5O7}{\bf (a)} is large and isomer Fig.~\ref{fig:spectra_Fe5O7}{\bf (b)} has a vibration at 450~\cmi{} which is not present in the experimental spectrum and lacks the experimental 375~\cmi{} vibration. Therefore, isomer Fig.~\ref{fig:spectra_Fe5O7}{\bf (a)} can be identified as the most probable ground state.  

\subsection{$\mathbf{Fe}_{6}\mathbf{O}_{8}^{+}$}
\begin{figure}[ht]
\includegraphics[width=8cm]{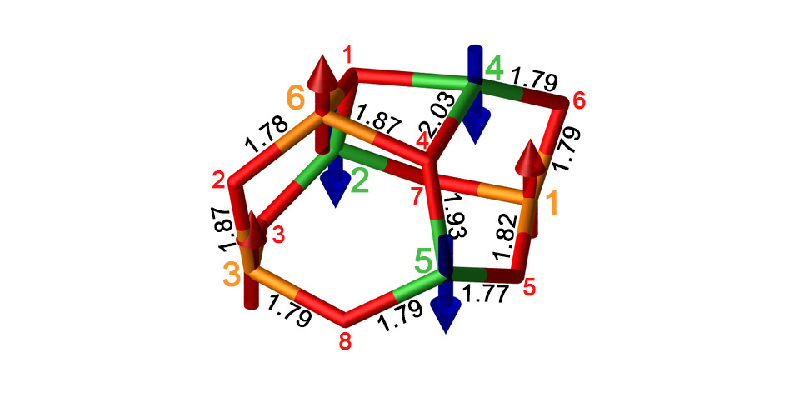}
\caption{(Color online) The cation \FeOp{6}{8} isomer lowest in energy. 
The cluster has $C_{s}$ symmetry. } 
\label{fig:Fe6O8p_struc}
\end{figure}

The isomer lowest in energy found for \FeOp{6}{8} is shown in Fig.~\ref{fig:Fe6O8p_struc} and has $C_{s}$ symmetry where the reflection plane is located through Fe atoms 1, 3, and 6. The magnetic moment of this isomer is 1~\muB{}. 

\begin{figure}[ht]
\includegraphics[width=8cm]{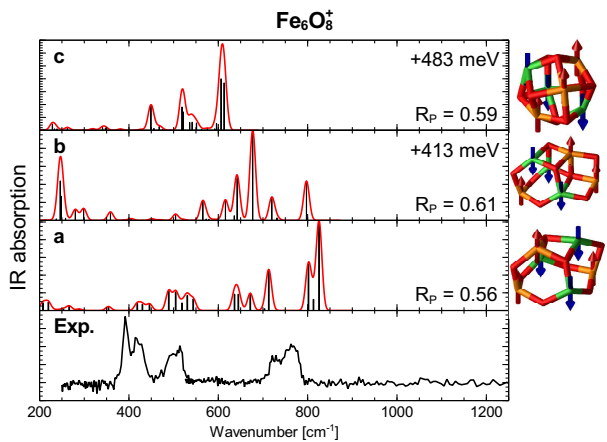}
\caption{(color online) The experimental and calculated vibration spectra of \FeOp{6}{8}. The isomer shown in {\bf (a)} is the lowest in energy. The reported energy differences include ZPVE. } 
\label{fig:spectra_Fe6O8}
\end{figure}

The second isomer low in energy is shown in Fig.~\ref{fig:spectra_Fe6O8}{\bf (b)}. In this isomer no symmetry is present. Compared to the lowest found isomer in Fig.~\ref{fig:spectra_Fe6O8}{\bf (a)} it is 413~meV higher in energy and also has a magnetic moment of 1~\muB{}. 

Figure~\ref{fig:spectra_Fe6O8}{\bf (c)} shows the third isomer, which is a distorted octahedral of Fe atoms in which the O atoms cap the Fe triangles. The structure is slightly distorted due to the AFM order between spins, which lead to slightly altered Fe-Fe distances. This isomer is 483~meV higher in energy than isomer Fig.~\ref{fig:spectra_Fe6O8}{\bf (a)}. 

Figure~\ref{fig:spectra_Fe6O8} also shows the corresponding vibration spectra of the mentioned isomers and the experimental spectrum. The experimental spectrum has vibrations at 392, 420, 500, 730 and 763~\cmi{}. Note that none of the provided isomers match the experimental vibration spectrum completely. This is also shown by the large-$R_{P}$ values of 0.56-0.61 for all calculated isomers. The isomer lowest in energy Fig.~\ref{fig:spectra_Fe6O8}{\bf (a)} is the best match since it also has vibrations at 420 and 500~\cmi{}, but the vibrations at 804 and 825 are considerably shifted with respect to 730 and 763~\cmi{}. Furthermore, the vibrations at 640, 671, and 713~\cmi{} are not present in the experimental spectrum. The vibration spectra shown in Figs.~\ref{fig:spectra_Fe6O8}{\bf (b)} and \ref{fig:spectra_Fe6O8}{\bf (c)} fit even worse. Therefore, we can not successfully identify the \FeOp{6}{8} structure. 

Note that our genetic algorithm implementation only uses geometry optimization at the DFT level. At cluster sizes of \FeOp{6}{8} and larger, preselection using empirical potentials instead of immediate geometry optimization using DFT might be more efficient in generating possible isomers.

\subsection{Electronic structure}
In the bulk, iron-oxide materials have many different crystal structures such as hematite, wustite, and magnetite with all corresponding different electronic structures. While in hematite only trivalent $\mathrm{Fe}^{3+}$ is present, the mixed valence state ($\mathrm{Fe}_{A}^{3+}[\mathrm{Fe}^{2+},\mathrm{Fe}^{3+}]_{B}\mathrm{O}_{4}$) in magnetite leads to interesting physical phenomena such as ferrimagnetic ordering between the sublattices $A$ and $B$ and the Verwey transition in which orbital ordering leads to a first-order phase transition in the electrical conductivity. \cite{Verwey1939,Walz2002}

In clusters, stoichiometries corresponding to both hematite (\FeO{4}{6}) and magnetite (\FeO{3}{4}, \FeO{6}{8}) and other combinations (\FeO{4}{5}, \FeO{5}{7}) occur. We therefore expect divalent and trivalent Fe cations to be present in the reported clusters. There is no unique method to determine the valence state in materials consisting of multiple types of elements. We therefore compare both the local magnetic moments and the local density of states (LDOS) for our cluster calculations with bulk magnetite results shown in Section~\ref{sec:magnetite}. 
Since the $\mathrm{Fe}^{2+}$ and $\mathrm{Fe}^{3+}$ features in the LDOS are very similar for different cluster sizes, we show the LDOS of \FeOp{4}{5} which contains both $\mathrm{Fe}^{2+}$ and $\mathrm{Fe}^{3+}$ in Fig.~\ref{fig:Fe4O5p_ldos}. The LDOS for other cluster sizes can be found in the Appendix. 

\begin{figure}[ht]
\includegraphics[width=8cm]{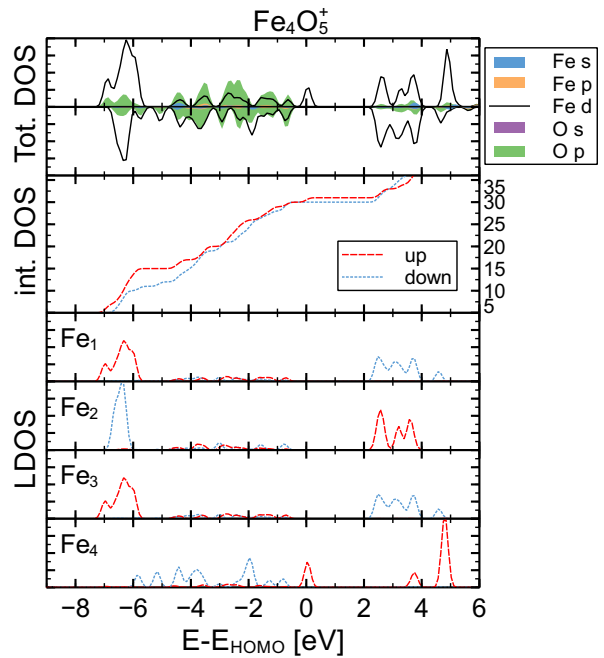}
\caption{(Color online) The total, integrated and local density of states of the Fe atoms for the \FeOp{4}{5} cluster. The trivalent Fe(1), Fe(2) and Fe(3) all show $3d$ levels at -6~eV and small hybridization with O. The divalent Fe(4), however, shows strong hybridization and a single level at $\text{E}_{\text{HOMO}}$. } 
\label{fig:Fe4O5p_ldos}
\end{figure}
\begin{table*}[hb]
\caption{The spin moment for \FeOp{x}{y} clusters. The atom numbers correspond to the atom numbers shown in Figures~\ref{fig:Fe3O4np_struc},\ref{fig:Fe4O5np_struc},\ref{fig:Fe4O6np_struc}, \ref{fig:Fe5O7np_struc}, and \ref{fig:Fe6O8p_struc}. The spin moment is calculated using atomic spheres of 1.3 and 0.82~\AAA{} for Fe and O, respectively. }
\label{tab:FeOElecstruc}
\begin{ruledtabular}
\begin{tabular}{lldddddddd}
Cluster & & \multicolumn{8}{l}{Spin moment [$\mu_{B}$]}\\
 & & 1 & 2 & 3 & 4 & 5 & 6 & 7 & 8 \\
\hline
\FeOp{3}{4} & Fe & -3.84 & 3.88 & 3.88 &  &  &  &  &  \\
            & O  & 0.56 & 0.00 & 0.00 & 0.22 &  &  &  &  \\
\FeOp{4}{5} & Fe & 3.89 & -3.84 & 3.89 & -3.40 &  &  &  &  \\
            & O  & -0.05 & 0.13 & 0.20 & -0.05 & 0.20 &  &  &  \\
\FeOp{4}{6} & Fe & -3.22 & 3.85 & 3.85 & -3.79 &  &  &  &  \\
            & O  & 0.01 & 0.54 & 0.01 & -0.25 & 0.00 & 0.00 &  &  \\        
\FeOp{5}{7} & Fe & 3.85 & 3.87 & 3.89 & -3.83 & -3.80 &  &  &  \\
            & O  & 0.01 & 0.10 & 0.03 & 0.51 & -0.09 & 0.05 & 0.12 &  \\
\FeOp{6}{8} & Fe & 3.80 &  -3.84 &   3.85 &  -3.47 &  -3.84 &   3.88 &  &  \\
            & O  & 0.01 & 0.51 & 0.01 & 0.01 &-0.10 & 0.17 &-0.10 & 0.01  \\
\end{tabular}
\end{ruledtabular}
\end{table*}

Table~\ref{tab:FeOElecstruc} shows the local spin moments of the clusters: \FeOp{3}{4}, \FeOp{4}{5}, \FeOp{4}{6}, \FeOp{5}{7} and \FeOp{6}{8}. For \FeOp{3}{4} all three Fe atoms have a similar spin moment within 0.04~\muB{}. A comparison with magnetite suggests all Fe atoms are trivalent. This agrees with an ionic bond model. Furthermore, this is confirmed by the integrated and local density of states shown in Appendix~\ref{sec:ldos}. The $3d$ peaks around -6~eV correspond to 15 electrons, indicating the hybridization between Fe and O is small. Note that, the central oxygen atoms O(4) and O(7) are partially spin polarized. 

For \FeOp{4}{5}, the spin moment of Fe(4) is 0.5~\muB{} lower than the other Fe atoms, indicating three trivalent and a single divalent atom. The difference is also in agreement with the magnetite results. The Fe(4) also breaks the $C_{2}$ symmetry as is shown in Fig.~\ref{fig:Fe4O5np_struc}. The local (LDOS) and integrated density of states are shown in Fig.~\ref{fig:Fe4O5p_ldos}. Note that all $\text{Fe}^{3+}$ have $3d$ peaks around $-6$~eV and small hybridization with O is present, similar to the \FeOp{3}{4} cluster. The LDOS of the divalent Fe(4) atom however shows strong hybridization with O and a single minority level at $\text{E}_{\text{HOMO}}$. 

\begin{figure}[hb]
\includegraphics[width=8cm]{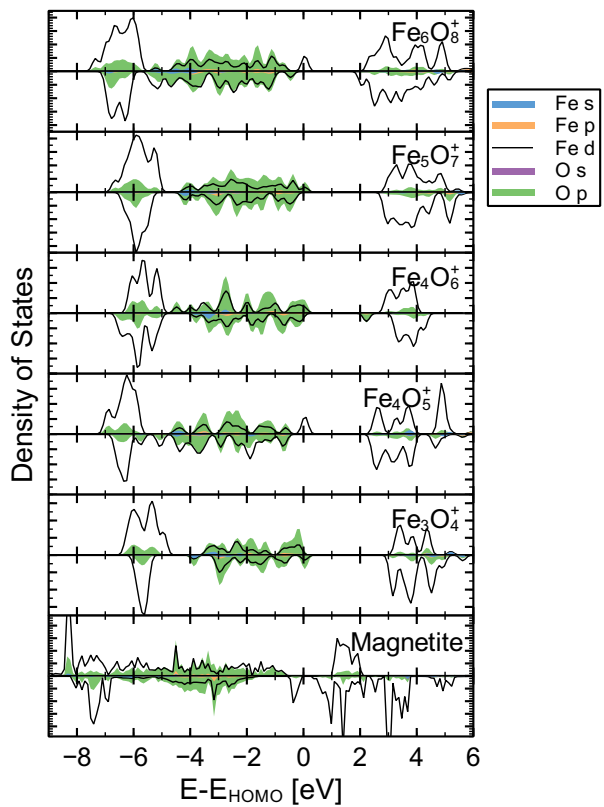}
\caption{(Color online) The density of states for \FeOp{x}{y} clusters. For these calculations a smearing of 0.15~eV was used for convenience of the reader. The HOMO level is located at 0~eV and the small occupation above the HOMO level is due to smearing. } 
\label{fig:dos}
\end{figure}

Whereas \FeO{4}{6} only contains trivalent Fe, \cite{Erlebach2014} for \FeOp{4}{6} this is no longer the case due to ionization. As can be seen from Table~\ref{tab:FeOElecstruc}, three trivalent Fe atoms are present, together with a single $\text{Fe}^{4+}$ atom. The spin moment is reduced with respect to $\text{Fe}^{3+}$, consistent with a higher oxidation state than $\text{Fe}^{3+}$.  

In \FeOp{5}{7}, only trivalent Fe atoms are present, consistent with an ionic model and the ionized state of the cluster. \FeOp{6}{8}, on the other hand, is again a mixed valence cluster where the magnetic moment of Fe(4) is 0.4~\muB{} lower than the other Fe atoms, indicating Fe(4) is divalent. This is also consistent with the LDOS shown in Appendix~\ref{sec:ldos}. 

Figure~\ref{fig:dos} shows the density of states for the different cationic clusters and magnetite. The calculated band gap of 0.2~eV in magnetite is considerably smaller than for the reported clusters: around 3~eV for \FeOp{3}{4} and slightly smaller for \FeOp{4}{5} and \FeOp{4}{6}. Furthermore, whereas magnetite has a $t_{2g}$ orbital of $\text{Fe}^{2+}$ just below the Fermi energy,\cite{Jeng2004} in the reported clusters \FeOp{4}{5} and \FeOp{6}{8} have a similar level due to a divalent Fe atom. Note that the $3d$ orbitals of $\text{Fe}^{3+}$ in the clusters are located around 5.5~eV below the HOMO level, which is 2~eV higher in energy compared to magnetite. 

\section{Conclusion}
In this work, we have studied the geometric, electronic and magnetic structure of \FeOp{x}{y} clusters using density functional theory. For \FeO{3}{4} we compared binding distances and electronic structure between the hybrid B3LYP functional, and different $U_{\text{eff}}$ in the PBE$+U$ formalism. We found the best match for $U_{\text{eff}}$ = 3~eV. Using the PBE$+U$ formalism and a genetic algorithm, many possible isomers were considered. For isomers low in energy, all different magnetic configurations were further geometrically optimized. Finally, for the cationic clusters we calculated the vibration spectra and compared them with experiments to identify the geometric structure of \FeOp{3}{4}, \FeOp{4}{5}, \FeOp{4}{6}, \FeOp{5}{7} and \FeOp{6}{8}. 
All cationic clusters with an even number of Fe atoms have a small magnetic moment of 1~\muB{} due to ionization. Furthermore, comparison with bulk magnetite reveals that \FeOp{4}{5}, \FeOp{4}{6} and \FeOp{6}{8} are mixed valence clusters. In contrast, in \FeOp{3}{4} and \FeOp{5}{7} all Fe are found to be trivalent. 

\section{Acknowledgements}
The work is supported by European Research Council (ERC) Advanced Grant No.
338957 FEMTO/NANO. 
\bibliographystyle{apsrev}
\bibliography{library}

\begin{thebibliography}{52}
\expandafter\ifx\csname natexlab\endcsname\relax\def\natexlab#1{#1}\fi
\expandafter\ifx\csname bibnamefont\endcsname\relax
  \def\bibnamefont#1{#1}\fi
\expandafter\ifx\csname bibfnamefont\endcsname\relax
  \def\bibfnamefont#1{#1}\fi
\expandafter\ifx\csname citenamefont\endcsname\relax
  \def\citenamefont#1{#1}\fi
\expandafter\ifx\csname url\endcsname\relax
  \def\url#1{\texttt{#1}}\fi
\expandafter\ifx\csname urlprefix\endcsname\relax\def\urlprefix{URL }\fi
\providecommand{\bibinfo}[2]{#2}
\providecommand{\eprint}[2][]{\url{#2}}

\bibitem[{\citenamefont{Verwey}(1939)}]{Verwey1939}
\bibinfo{author}{\bibfnamefont{E.~J.~W.} \bibnamefont{Verwey}},
  \bibinfo{journal}{Nature} \textbf{\bibinfo{volume}{144}},
  \bibinfo{pages}{327} (\bibinfo{year}{1939}).

\bibitem[{\citenamefont{Walz}(2002)}]{Walz2002}
\bibinfo{author}{\bibfnamefont{F.}~\bibnamefont{Walz}},
  \bibinfo{journal}{J. Phys.: Condens. Matter}
  \textbf{\bibinfo{volume}{14}}, \bibinfo{pages}{R285} (\bibinfo{year}{2002}).

\bibitem[{\citenamefont{Andrews et~al.}(1996)\citenamefont{Andrews, Chertihin,
  Ricca, and Bauschlicher}}]{Andrews1996}
\bibinfo{author}{\bibfnamefont{L.}~\bibnamefont{Andrews}},
  \bibinfo{author}{\bibfnamefont{G.~V.} \bibnamefont{Chertihin}},
  \bibinfo{author}{\bibfnamefont{A.}~\bibnamefont{Ricca}}, \bibnamefont{and}
  \bibinfo{author}{\bibfnamefont{C.~W.} \bibnamefont{Bauschlicher}},
  \bibinfo{journal}{J. Am. Chem. Soc.}
  \textbf{\bibinfo{volume}{118}}, \bibinfo{pages}{467} (\bibinfo{year}{1996}).

\bibitem[{\citenamefont{Chertihin et~al.}(1996)\citenamefont{Chertihin, Saffel,
  Yustein, Andrews, Neurock, Ricca, and Bauschlicher}}]{Chertihin1996}
\bibinfo{author}{\bibfnamefont{G.~V.} \bibnamefont{Chertihin}},
  \bibinfo{author}{\bibfnamefont{W.}~\bibnamefont{Saffel}},
  \bibinfo{author}{\bibfnamefont{J.~T.} \bibnamefont{Yustein}},
  \bibinfo{author}{\bibfnamefont{L.}~\bibnamefont{Andrews}},
  \bibinfo{author}{\bibfnamefont{M.}~\bibnamefont{Neurock}},
  \bibinfo{author}{\bibfnamefont{A.}~\bibnamefont{Ricca}}, \bibnamefont{and}
  \bibinfo{author}{\bibfnamefont{C.~W.} \bibnamefont{Bauschlicher}},
  \bibinfo{journal}{J. Phys. Chem.}
  \textbf{\bibinfo{volume}{100}}, \bibinfo{pages}{5261} (\bibinfo{year}{1996}).

\bibitem[{\citenamefont{Laurent et~al.}(2008)\citenamefont{Laurent, Forge,
  Port, Roch, Robic, {Vander Elst}, and Muller}}]{Laurent2008}
\bibinfo{author}{\bibfnamefont{S.}~\bibnamefont{Laurent}},
  \bibinfo{author}{\bibfnamefont{D.}~\bibnamefont{Forge}},
  \bibinfo{author}{\bibfnamefont{M.}~\bibnamefont{Port}},
  \bibinfo{author}{\bibfnamefont{A.}~\bibnamefont{Roch}},
  \bibinfo{author}{\bibfnamefont{C.}~\bibnamefont{Robic}},
  \bibinfo{author}{\bibfnamefont{L.}~\bibnamefont{{Vander Elst}}},
  \bibnamefont{and} \bibinfo{author}{\bibfnamefont{R.~N.}
  \bibnamefont{Muller}}, \bibinfo{journal}{Chem. Rev.}
  \textbf{\bibinfo{volume}{108}}, \bibinfo{pages}{2064} (\bibinfo{year}{2008}).

\bibitem[{\citenamefont{Reilly et~al.}(2007)\citenamefont{Reilly, Reveles,
  Johnson, Khanna, and Castleman}}]{Reilly2007}
\bibinfo{author}{\bibfnamefont{N.~M.} \bibnamefont{Reilly}},
  \bibinfo{author}{\bibfnamefont{J.~U.} \bibnamefont{Reveles}},
  \bibinfo{author}{\bibfnamefont{G.~E.} \bibnamefont{Johnson}},
  \bibinfo{author}{\bibfnamefont{S.~N.} \bibnamefont{Khanna}},
  \bibnamefont{and} \bibinfo{author}{\bibfnamefont{A.~W.}
  \bibnamefont{Castleman}}, \bibinfo{journal}{J. Phys. Chem. A} 
  \textbf{\bibinfo{volume}{111}}, \bibinfo{pages}{4158}
  (\bibinfo{year}{2007}).

\bibitem[{\citenamefont{Wang et~al.}(1996)\citenamefont{Wang, Wu, and
  Desai}}]{Wang1996}
\bibinfo{author}{\bibfnamefont{L.~S.} \bibnamefont{Wang}},
  \bibinfo{author}{\bibfnamefont{H.}~\bibnamefont{Wu}}, \bibnamefont{and}
  \bibinfo{author}{\bibfnamefont{S.~R.} \bibnamefont{Desai}},
  \bibinfo{journal}{Phys. Rev. Lett.} \textbf{\bibinfo{volume}{76}},
  \bibinfo{pages}{4853} (\bibinfo{year}{1996}).

\bibitem[{\citenamefont{Schr\"{o}der et~al.}(2000)\citenamefont{Schr\"{o}der,
  Jackson, and Schwarz}}]{Schroder2000}
\bibinfo{author}{\bibfnamefont{D.}~\bibnamefont{Schr\"{o}der}},
  \bibinfo{author}{\bibfnamefont{P.}~\bibnamefont{Jackson}}, \bibnamefont{and}
  \bibinfo{author}{\bibfnamefont{H.}~\bibnamefont{Schwarz}},
  \bibinfo{journal}{Eur. J. Inorg. Chem.}
  \textbf{\bibinfo{volume}{2000}}, \bibinfo{pages}{1171}
  (\bibinfo{year}{2000}).

\bibitem[{\citenamefont{Erlebach et~al.}(2015)\citenamefont{Erlebach, Kurland,
  Grabow, M\"{u}ller, and Sierka}}]{Erlebach2015}
\bibinfo{author}{\bibfnamefont{A.}~\bibnamefont{Erlebach}},
  \bibinfo{author}{\bibfnamefont{H.~D.} \bibnamefont{Kurland}},
  \bibinfo{author}{\bibfnamefont{J.}~\bibnamefont{Grabow}},
  \bibinfo{author}{\bibfnamefont{F.~A.} \bibnamefont{M\"{u}ller}},
  \bibnamefont{and} \bibinfo{author}{\bibfnamefont{M.}~\bibnamefont{Sierka}},
  \bibinfo{journal}{Nanoscale} \textbf{\bibinfo{volume}{7}},
  \bibinfo{pages}{2960} (\bibinfo{year}{2015}).

\bibitem[{\citenamefont{Reddy et~al.}(2004)\citenamefont{Reddy, Rasouli,
  Hajaligol, and Khanna}}]{Reddy2004}
\bibinfo{author}{\bibfnamefont{B.~V.} \bibnamefont{Reddy}},
  \bibinfo{author}{\bibfnamefont{F.}~\bibnamefont{Rasouli}},
  \bibinfo{author}{\bibfnamefont{M.~R.} \bibnamefont{Hajaligol}},
  \bibnamefont{and} \bibinfo{author}{\bibfnamefont{S.~N.}
  \bibnamefont{Khanna}}, \bibinfo{journal}{Fuel} \textbf{\bibinfo{volume}{83}},
  \bibinfo{pages}{1537} (\bibinfo{year}{2004}).

\bibitem[{\citenamefont{Reddy and Khanna}(2004)}]{Reddy2004a}
\bibinfo{author}{\bibfnamefont{B.~V.} \bibnamefont{Reddy}} \bibnamefont{and}
  \bibinfo{author}{\bibfnamefont{S.~N.} \bibnamefont{Khanna}},
  \bibinfo{journal}{Phys. Rev. Lett.} \textbf{\bibinfo{volume}{93}},
  \bibinfo{pages}{068301} (\bibinfo{year}{2004}).

\bibitem[{\citenamefont{Fiedler et~al.}(1994)\citenamefont{Fiedler, Schroeder,
  Shaik, and Schwarz}}]{Fiedler1994}
\bibinfo{author}{\bibfnamefont{A.}~\bibnamefont{Fiedler}},
  \bibinfo{author}{\bibfnamefont{D.}~\bibnamefont{Schroeder}},
  \bibinfo{author}{\bibfnamefont{S.}~\bibnamefont{Shaik}}, \bibnamefont{and}
  \bibinfo{author}{\bibfnamefont{H.}~\bibnamefont{Schwarz}},
  \bibinfo{journal}{J. Am. Chem. Soc.}
  \textbf{\bibinfo{volume}{116}}, \bibinfo{pages}{10734}
  (\bibinfo{year}{1994}).

\bibitem[{\citenamefont{Ohshimo et~al.}(2014)\citenamefont{Ohshimo, Komukai,
  Moriyama, and Misaizu}}]{Ohshimo2014}
\bibinfo{author}{\bibfnamefont{K.}~\bibnamefont{Ohshimo}},
  \bibinfo{author}{\bibfnamefont{T.}~\bibnamefont{Komukai}},
  \bibinfo{author}{\bibfnamefont{R.}~\bibnamefont{Moriyama}}, \bibnamefont{and}
  \bibinfo{author}{\bibfnamefont{F.}~\bibnamefont{Misaizu}},
  \bibinfo{journal}{J. Phys. Chem. A}
  \textbf{\bibinfo{volume}{118}}, \bibinfo{pages}{3899} (\bibinfo{year}{2014}).

\bibitem[{\citenamefont{Yin et~al.}(2009)\citenamefont{Yin, Xue, Ding, Wang,
  He, and Ge}}]{Yin2009}
\bibinfo{author}{\bibfnamefont{S.}~\bibnamefont{Yin}},
  \bibinfo{author}{\bibfnamefont{W.}~\bibnamefont{Xue}},
  \bibinfo{author}{\bibfnamefont{X.~L.} \bibnamefont{Ding}},
  \bibinfo{author}{\bibfnamefont{W.~G.} \bibnamefont{Wang}},
  \bibinfo{author}{\bibfnamefont{S.~G.} \bibnamefont{He}}, \bibnamefont{and}
  \bibinfo{author}{\bibfnamefont{M.~F.} \bibnamefont{Ge}},
  \bibinfo{journal}{Int. J. Mass Spectrom.}
  \textbf{\bibinfo{volume}{281}}, \bibinfo{pages}{72} (\bibinfo{year}{2009}).

\bibitem[{\citenamefont{Palot\'{a}s et~al.}(2010)\citenamefont{Palot\'{a}s,
  Andriotis, and Lappas}}]{Palotas2010}
\bibinfo{author}{\bibfnamefont{K.}~\bibnamefont{Palot\'{a}s}},
  \bibinfo{author}{\bibfnamefont{A.~N.} \bibnamefont{Andriotis}},
  \bibnamefont{and} \bibinfo{author}{\bibfnamefont{A.}~\bibnamefont{Lappas}},
  \bibinfo{journal}{Phys. Rev. B} \textbf{\bibinfo{volume}{81}},
  \bibinfo{pages}{075403} (\bibinfo{year}{2010}).

\bibitem[{\citenamefont{Sun et~al.}(2000{\natexlab{a}})\citenamefont{Sun, Wang,
  Parlinski, Yu, Hashi, Gong, and Kawazoe}}]{Sun2000}
\bibinfo{author}{\bibfnamefont{Q.}~\bibnamefont{Sun}},
  \bibinfo{author}{\bibfnamefont{Q.}~\bibnamefont{Wang}},
  \bibinfo{author}{\bibfnamefont{K.}~\bibnamefont{Parlinski}},
  \bibinfo{author}{\bibfnamefont{J.~Z.} \bibnamefont{Yu}},
  \bibinfo{author}{\bibfnamefont{Y.}~\bibnamefont{Hashi}},
  \bibinfo{author}{\bibfnamefont{X.~G.} \bibnamefont{Gong}}, \bibnamefont{and}
  \bibinfo{author}{\bibfnamefont{Y.}~\bibnamefont{Kawazoe}},
  \bibinfo{journal}{Phys. Rev. B} \textbf{\bibinfo{volume}{61}},
  \bibinfo{pages}{5781} (\bibinfo{year}{2000}{\natexlab{a}}).

\bibitem[{\citenamefont{Wang et~al.}(1999)\citenamefont{Wang, Sun, Sakurai, Yu,
  Gu, Sumiyama, and Kawazoe}}]{Wang1999}
\bibinfo{author}{\bibfnamefont{Q.}~\bibnamefont{Wang}},
  \bibinfo{author}{\bibfnamefont{Q.}~\bibnamefont{Sun}},
  \bibinfo{author}{\bibfnamefont{M.}~\bibnamefont{Sakurai}},
  \bibinfo{author}{\bibfnamefont{J.~Z.} \bibnamefont{Yu}},
  \bibinfo{author}{\bibfnamefont{B.~L.} \bibnamefont{Gu}},
  \bibinfo{author}{\bibfnamefont{K.}~\bibnamefont{Sumiyama}}, \bibnamefont{and}
  \bibinfo{author}{\bibfnamefont{Y.}~\bibnamefont{Kawazoe}},
  \bibinfo{journal}{Phys. Rev. B} \textbf{\bibinfo{volume}{59}},
  \bibinfo{pages}{12672} (\bibinfo{year}{1999}).

\bibitem[{\citenamefont{Kortus and Pederson}(2000)}]{Kortus2000}
\bibinfo{author}{\bibfnamefont{J.}~\bibnamefont{Kortus}} \bibnamefont{and}
  \bibinfo{author}{\bibfnamefont{M.~R.} \bibnamefont{Pederson}},
  \bibinfo{journal}{Phys. Rev. B} \textbf{\bibinfo{volume}{62}},
  \bibinfo{pages}{5755} (\bibinfo{year}{2000}).

\bibitem[{\citenamefont{Sun et~al.}(2007)\citenamefont{Sun, Reddy, Marquez,
  Jena, Gonzalez, and Wang}}]{Sun2007}
\bibinfo{author}{\bibfnamefont{Q.}~\bibnamefont{Sun}},
  \bibinfo{author}{\bibfnamefont{B.~V.} \bibnamefont{Reddy}},
  \bibinfo{author}{\bibfnamefont{M.}~\bibnamefont{Marquez}},
  \bibinfo{author}{\bibfnamefont{P.}~\bibnamefont{Jena}},
  \bibinfo{author}{\bibfnamefont{C.}~\bibnamefont{Gonzalez}}, \bibnamefont{and}
  \bibinfo{author}{\bibfnamefont{Q.}~\bibnamefont{Wang}},
  \bibinfo{journal}{J. Phys. Chem. C}
  \textbf{\bibinfo{volume}{111}}, \bibinfo{pages}{4159} (\bibinfo{year}{2007}).

\bibitem[{\citenamefont{L\'{o}pez et~al.}(2009)\citenamefont{L\'{o}pez, Romero,
  Mej\'{\i}a-L\'{o}pez, Mazo-Zuluaga, and Restrepo}}]{Lopez2009}
\bibinfo{author}{\bibfnamefont{S.}~\bibnamefont{L\'{o}pez}},
  \bibinfo{author}{\bibfnamefont{A.~H.}~\bibnamefont{Romero}},
  \bibinfo{author}{\bibfnamefont{J.}~\bibnamefont{Mej\'{\i}a-L\'{o}pez}},
  \bibinfo{author}{\bibfnamefont{J.}~\bibnamefont{Mazo-Zuluaga}},
  \bibnamefont{and} \bibinfo{author}{\bibfnamefont{J.}~\bibnamefont{Restrepo}},
  \bibinfo{journal}{Phys. Rev. B} \textbf{\bibinfo{volume}{80}},
  \bibinfo{pages}{085107} (\bibinfo{year}{2009}).

\bibitem[{\citenamefont{Ding et~al.}(2009)\citenamefont{Ding, Xue, Ma, Wang,
  and He}}]{Ding2009}
\bibinfo{author}{\bibfnamefont{X.~L.} \bibnamefont{Ding}},
  \bibinfo{author}{\bibfnamefont{W.}~\bibnamefont{Xue}},
  \bibinfo{author}{\bibfnamefont{Y.~P.} \bibnamefont{Ma}},
  \bibinfo{author}{\bibfnamefont{Z.~C.} \bibnamefont{Wang}}, \bibnamefont{and}
  \bibinfo{author}{\bibfnamefont{S.~G.} \bibnamefont{He}},
  \bibinfo{journal}{J. Chem. Phys.}
  \textbf{\bibinfo{volume}{130}}, \bibinfo{pages}{014303}
  (\bibinfo{year}{2009}).

\bibitem[{\citenamefont{Jones et~al.}(2005)\citenamefont{Jones, Reddy, Rasouli,
  and Khanna}}]{Jones2005}
\bibinfo{author}{\bibfnamefont{N.~O.} \bibnamefont{Jones}},
  \bibinfo{author}{\bibfnamefont{B.~V.} \bibnamefont{Reddy}},
  \bibinfo{author}{\bibfnamefont{F.}~\bibnamefont{Rasouli}}, \bibnamefont{and}
  \bibinfo{author}{\bibfnamefont{S.~N.} \bibnamefont{Khanna}},
  \bibinfo{journal}{Phys. Rev. B} \textbf{\bibinfo{volume}{72}},
  \bibinfo{pages}{165411} (\bibinfo{year}{2005}).

\bibitem[{\citenamefont{Shiroishi et~al.}(2003)\citenamefont{Shiroishi, Oda,
  Hamada, and Fujima}}]{Shiroishi2003}
\bibinfo{author}{\bibfnamefont{H.}~\bibnamefont{Shiroishi}},
  \bibinfo{author}{\bibfnamefont{T.}~\bibnamefont{Oda}},
  \bibinfo{author}{\bibfnamefont{I.}~\bibnamefont{Hamada}}, \bibnamefont{and}
  \bibinfo{author}{\bibfnamefont{N.}~\bibnamefont{Fujima}},
  \bibinfo{journal}{Eur. Phys. J. D} \textbf{\bibinfo{volume}{24}}, \bibinfo{pages}{85}
  (\bibinfo{year}{2003}).

\bibitem[{\citenamefont{Sun et~al.}(2000{\natexlab{b}})\citenamefont{Sun,
  Sakurai, Wang, Yu, Wang, Sumiyama, and Kawazoe}}]{Sun2000a}
\bibinfo{author}{\bibfnamefont{Q.}~\bibnamefont{Sun}},
  \bibinfo{author}{\bibfnamefont{M.}~\bibnamefont{Sakurai}},
  \bibinfo{author}{\bibfnamefont{Q.}~\bibnamefont{Wang}},
  \bibinfo{author}{\bibfnamefont{J.~Z.} \bibnamefont{Yu}},
  \bibinfo{author}{\bibfnamefont{G.~H.} \bibnamefont{Wang}},
  \bibinfo{author}{\bibfnamefont{K.}~\bibnamefont{Sumiyama}}, \bibnamefont{and}
  \bibinfo{author}{\bibfnamefont{Y.}~\bibnamefont{Kawazoe}},
  \bibinfo{journal}{Phys. Rev. B} \textbf{\bibinfo{volume}{62}},
  \bibinfo{pages}{8500} (\bibinfo{year}{2000}{\natexlab{b}}).

\bibitem[{\citenamefont{Cao et~al.}(1998)\citenamefont{Cao, Duran, and
  Sol\`{a}}}]{Cao1998}
\bibinfo{author}{\bibfnamefont{Z.}~\bibnamefont{Cao}},
  \bibinfo{author}{\bibfnamefont{M.}~\bibnamefont{Duran}}, \bibnamefont{and}
  \bibinfo{author}{\bibfnamefont{M.}~\bibnamefont{Sol\`{a}}},
  \bibinfo{journal}{J. Chem. Soc., Faraday Trans.}
  \textbf{\bibinfo{volume}{94}}, \bibinfo{pages}{2877} (\bibinfo{year}{1998}).

\bibitem[{\citenamefont{Erlebach et~al.}(2014)\citenamefont{Erlebach, H\"{u}hn,
  Jana, and Sierka}}]{Erlebach2014}
\bibinfo{author}{\bibfnamefont{A.}~\bibnamefont{Erlebach}},
  \bibinfo{author}{\bibfnamefont{C.}~\bibnamefont{H\"{u}hn}},
  \bibinfo{author}{\bibfnamefont{R.}~\bibnamefont{Jana}}, \bibnamefont{and}
  \bibinfo{author}{\bibfnamefont{M.}~\bibnamefont{Sierka}},
  \bibinfo{journal}{Phys. Chem. Chem. Phys.} \textbf{\bibinfo{volume}{16}},
  \bibinfo{pages}{26421} (\bibinfo{year}{2014}).

\bibitem[{\citenamefont{Johnston}(2003)}]{Johnston2003}
\bibinfo{author}{\bibfnamefont{R.~L.} \bibnamefont{Johnston}},
  \bibinfo{journal}{Dalton Trans.} \textbf{\bibinfo{volume}{2003}},
  \bibinfo{pages}{4193} (\bibinfo{year}{2003}).

\bibitem[{\citenamefont{Haertelt et~al.}(2012)\citenamefont{Haertelt, Fielicke,
  Meijer, Kwapien, Sierka, and Sauer}}]{Haertelt2012}
\bibinfo{author}{\bibfnamefont{M.}~\bibnamefont{Haertelt}},
  \bibinfo{author}{\bibfnamefont{A.}~\bibnamefont{Fielicke}},
  \bibinfo{author}{\bibfnamefont{G.}~\bibnamefont{Meijer}},
  \bibinfo{author}{\bibfnamefont{K.}~\bibnamefont{Kwapien}},
  \bibinfo{author}{\bibfnamefont{M.}~\bibnamefont{Sierka}}, \bibnamefont{and}
  \bibinfo{author}{\bibfnamefont{J.}~\bibnamefont{Sauer}},
  \bibinfo{journal}{Phys. Chem. Chem. Phys.}
  \textbf{\bibinfo{volume}{14}}, \bibinfo{pages}{2849} (\bibinfo{year}{2012}).

\bibitem[{\citenamefont{Zhai et~al.}(2007)\citenamefont{Zhai, D\"{o}bler,
  Sauer, and Wang}}]{Zhai2007a}
\bibinfo{author}{\bibfnamefont{H.~J.} \bibnamefont{Zhai}},
  \bibinfo{author}{\bibfnamefont{J.}~\bibnamefont{D\"{o}bler}},
  \bibinfo{author}{\bibfnamefont{J.}~\bibnamefont{Sauer}}, \bibnamefont{and}
  \bibinfo{author}{\bibfnamefont{L.~S.} \bibnamefont{Wang}},
  \bibinfo{journal}{J. Am. Chem. Soc.}
  \textbf{\bibinfo{volume}{129}}, \bibinfo{pages}{13270}
  (\bibinfo{year}{2007}).

\bibitem[{\citenamefont{Kirilyuk et~al.}(2010)\citenamefont{Kirilyuk, Fielicke,
  Demyk, von Helden, Meijer, and Rasing}}]{Kirilyuk2010}
\bibinfo{author}{\bibfnamefont{A.}~\bibnamefont{Kirilyuk}},
  \bibinfo{author}{\bibfnamefont{A.}~\bibnamefont{Fielicke}},
  \bibinfo{author}{\bibfnamefont{K.}~\bibnamefont{Demyk}},
  \bibinfo{author}{\bibfnamefont{G.}~\bibnamefont{von Helden}},
  \bibinfo{author}{\bibfnamefont{G.}~\bibnamefont{Meijer}}, \bibnamefont{and}
  \bibinfo{author}{\bibfnamefont{T.}~\bibnamefont{Rasing}},
  \bibinfo{journal}{Phys. Rev. B} \textbf{\bibinfo{volume}{82}},
  \bibinfo{pages}{020405} (\bibinfo{year}{2010}).

\bibitem[{\citenamefont{Kresse and Furthm\"{u}ller}(1996)}]{Kresse1996}
\bibinfo{author}{\bibfnamefont{G.}~\bibnamefont{Kresse}} \bibnamefont{and}
  \bibinfo{author}{\bibfnamefont{J.}~\bibnamefont{Furthm\"{u}ller}},
  \bibinfo{journal}{Phys. Rev. B} \textbf{\bibinfo{volume}{54}},
  \bibinfo{pages}{11169} (\bibinfo{year}{1996}).

\bibitem[{\citenamefont{Bl\"{o}chl}(1994)}]{Blochl1994}
\bibinfo{author}{\bibfnamefont{P.~E.} \bibnamefont{Bl\"{o}chl}},
  \bibinfo{journal}{Phys. Rev. B} \textbf{\bibinfo{volume}{50}},
  \bibinfo{pages}{17953} (\bibinfo{year}{1994}).

\bibitem[{\citenamefont{Kresse and Joubert}(1999)}]{Kresse1999}
\bibinfo{author}{\bibfnamefont{G.}~\bibnamefont{Kresse}} \bibnamefont{and}
  \bibinfo{author}{\bibfnamefont{D.}~\bibnamefont{Joubert}},
  \bibinfo{journal}{Phys. Rev. B} \textbf{\bibinfo{volume}{59}},
  \bibinfo{pages}{1758} (\bibinfo{year}{1999}).

\bibitem[{\citenamefont{Perdew et~al.}(1996)\citenamefont{Perdew, Burke, and
  Ernzerhof}}]{Perdew1996}
\bibinfo{author}{\bibfnamefont{J.~P.} \bibnamefont{Perdew}},
  \bibinfo{author}{\bibfnamefont{K.}~\bibnamefont{Burke}}, \bibnamefont{and}
  \bibinfo{author}{\bibfnamefont{M.}~\bibnamefont{Ernzerhof}},
  \bibinfo{journal}{Phys. Rev. Lett.} \textbf{\bibinfo{volume}{77}},
  \bibinfo{pages}{3865} (\bibinfo{year}{1996}).

\bibitem[{\citenamefont{Anisimov and Izyumov}(2002)}]{Anisimov2002}
\bibinfo{author}{\bibfnamefont{V.~I.} \bibnamefont{Anisimov}} \bibnamefont{and}
  \bibinfo{author}{\bibfnamefont{Y.}~\bibnamefont{Izyumov}},
  \emph{\bibinfo{title}{{Electronic Structure of Strongly Correlated
  Materials}}} (Springer-Verlag Berlin Heidelberg, \bibinfo{year}{2010}).

\bibitem[{\citenamefont{Anisimov et~al.}(1997)\citenamefont{Anisimov,
  Aryasetiawan, and Lichtenstein}}]{Anisimov1997}
\bibinfo{author}{\bibfnamefont{V.~I.} \bibnamefont{Anisimov}},
  \bibinfo{author}{\bibfnamefont{F.}~\bibnamefont{Aryasetiawan}},
  \bibnamefont{and} \bibinfo{author}{\bibfnamefont{A.~I.}
  \bibnamefont{Lichtenstein}}, \bibinfo{journal}{J. Phys.: Condens. Matter} 
  \textbf{\bibinfo{volume}{9}}, \bibinfo{pages}{767}
  (\bibinfo{year}{1997}).

\bibitem[{\citenamefont{Burow et~al.}(2011)\citenamefont{Burow, Wende, Sierka,
  Włodarczyk, Sauer, Claes, Jiang, Meijer, Lievens, and Asmis}}]{Burow2011}
\bibinfo{author}{\bibfnamefont{A.~M.} \bibnamefont{Burow}},
  \bibinfo{author}{\bibfnamefont{T.}~\bibnamefont{Wende}},
  \bibinfo{author}{\bibfnamefont{M.}~\bibnamefont{Sierka}},
  \bibinfo{author}{\bibfnamefont{R.}~\bibnamefont{Włodarczyk}},
  \bibinfo{author}{\bibfnamefont{J.}~\bibnamefont{Sauer}},
  \bibinfo{author}{\bibfnamefont{P.}~\bibnamefont{Claes}},
  \bibinfo{author}{\bibfnamefont{L.}~\bibnamefont{Jiang}},
  \bibinfo{author}{\bibfnamefont{G.}~\bibnamefont{Meijer}},
  \bibinfo{author}{\bibfnamefont{P.}~\bibnamefont{Lievens}}, \bibnamefont{and}
  \bibinfo{author}{\bibfnamefont{K.~R.} \bibnamefont{Asmis}},
  \bibinfo{journal}{Phys. Chem. Chem. Phys.}
  \textbf{\bibinfo{volume}{13}}, \bibinfo{pages}{19393} (\bibinfo{year}{2011}).

\bibitem[{\citenamefont{Dudarev et~al.}(1998)\citenamefont{Dudarev, Botton,
  Savrasov, Humphreys, and Sutton}}]{Dudarev1998}
\bibinfo{author}{\bibfnamefont{S.~L.} \bibnamefont{Dudarev}},
  \bibinfo{author}{\bibfnamefont{G.~A.} \bibnamefont{Botton}},
  \bibinfo{author}{\bibfnamefont{S.~Y.} \bibnamefont{Savrasov}},
  \bibinfo{author}{\bibfnamefont{C.~J.} \bibnamefont{Humphreys}},
  \bibnamefont{and} \bibinfo{author}{\bibfnamefont{A.~P.}
  \bibnamefont{Sutton}}, \bibinfo{journal}{Phys. Rev. B}
  \textbf{\bibinfo{volume}{57}}, \bibinfo{pages}{1505} (\bibinfo{year}{1998}).

\bibitem[{\citenamefont{Jeng et~al.}(2004)\citenamefont{Jeng, Guo, and
  Huang}}]{Jeng2004}
\bibinfo{author}{\bibfnamefont{H.~T.} \bibnamefont{Jeng}},
  \bibinfo{author}{\bibfnamefont{G.~Y.} \bibnamefont{Guo}}, \bibnamefont{and}
  \bibinfo{author}{\bibfnamefont{D.~J.} \bibnamefont{Huang}},
  \bibinfo{journal}{Phys. Rev. Lett.} \textbf{\bibinfo{volume}{93}},
  \bibinfo{pages}{156403} (\bibinfo{year}{2004}).

\bibitem[{\citenamefont{Becke}(1993)}]{Becke1993}
\bibinfo{author}{\bibfnamefont{A.~D.} \bibnamefont{Becke}},
  \bibinfo{journal}{J. Chem. Phys.}
  \textbf{\bibinfo{volume}{98}}, \bibinfo{pages}{1372} (\bibinfo{year}{1993}).

\bibitem[{\citenamefont{Makov and Payne}(1995)}]{Makov1995}
\bibinfo{author}{\bibfnamefont{G.}~\bibnamefont{Makov}} \bibnamefont{and}
  \bibinfo{author}{\bibfnamefont{M.~C.} \bibnamefont{Payne}},
  \bibinfo{journal}{Phys. Rev. B} \textbf{\bibinfo{volume}{51}},
  \bibinfo{pages}{4014} (\bibinfo{year}{1995}).

\bibitem[{\citenamefont{Neugebauer and Scheffler}(1992)}]{Neugebauer1992}
\bibinfo{author}{\bibfnamefont{J.}~\bibnamefont{Neugebauer}} \bibnamefont{and}
  \bibinfo{author}{\bibfnamefont{M.}~\bibnamefont{Scheffler}},
  \bibinfo{journal}{Phys. Rev. B} \textbf{\bibinfo{volume}{46}},
  \bibinfo{pages}{16067} (\bibinfo{year}{1992}).

\bibitem[{\citenamefont{Fan and Ziegler}(1992)}]{Fan1992}
\bibinfo{author}{\bibfnamefont{L.}~\bibnamefont{Fan}} \bibnamefont{and}
  \bibinfo{author}{\bibfnamefont{T.}~\bibnamefont{Ziegler}},
  \bibinfo{journal}{J. Chem. Phys.}
  \textbf{\bibinfo{volume}{96}}, \bibinfo{pages}{9005} (\bibinfo{year}{1992}).

\bibitem[{\citenamefont{Porezag and Pederson}(1996)}]{Porezag1996}
\bibinfo{author}{\bibfnamefont{D.}~\bibnamefont{Porezag}} \bibnamefont{and}
  \bibinfo{author}{\bibfnamefont{M.~R.} \bibnamefont{Pederson}},
  \bibinfo{journal}{Phys. Rev. B} \textbf{\bibinfo{volume}{54}},
  \bibinfo{pages}{7830} (\bibinfo{year}{1996}).

\bibitem[{\citenamefont{Pendry}(1980)}]{Pendry2000}
\bibinfo{author}{\bibfnamefont{J.~B.} \bibnamefont{Pendry}},
  \bibinfo{journal}{J. Phys. C}
  \textbf{\bibinfo{volume}{13}}, \bibinfo{pages}{937} (\bibinfo{year}{1980}).

\bibitem[{\citenamefont{Rossi et~al.}(2010)\citenamefont{Rossi, Blum, Kupser,
  {Von Helden}, Bierau, Pagel, Meijer, and Scheffler}}]{Rossi2010}
\bibinfo{author}{\bibfnamefont{M.}~\bibnamefont{Rossi}},
  \bibinfo{author}{\bibfnamefont{V.}~\bibnamefont{Blum}},
  \bibinfo{author}{\bibfnamefont{P.}~\bibnamefont{Kupser}},
  \bibinfo{author}{\bibfnamefont{G.}~\bibnamefont{{Von Helden}}},
  \bibinfo{author}{\bibfnamefont{F.}~\bibnamefont{Bierau}},
  \bibinfo{author}{\bibfnamefont{K.}~\bibnamefont{Pagel}},
  \bibinfo{author}{\bibfnamefont{G.}~\bibnamefont{Meijer}}, \bibnamefont{and}
  \bibinfo{author}{\bibfnamefont{M.}~\bibnamefont{Scheffler}},
  \bibinfo{journal}{J. Phys. Chem. Lett.}
  \textbf{\bibinfo{volume}{1}}, \bibinfo{pages}{3465} (\bibinfo{year}{2010}).

\bibitem[{\citenamefont{Lichtenstein et~al.}(1995)\citenamefont{Lichtenstein,
  Anisimov, and Zaanen}}]{Lichtenstein1995}
\bibinfo{author}{\bibfnamefont{A.~I.} \bibnamefont{Liechtenstein}},
  \bibinfo{author}{\bibfnamefont{V.~I.} \bibnamefont{Anisimov}},
  \bibnamefont{and} \bibinfo{author}{\bibfnamefont{J.}~\bibnamefont{Zaanen}},
  \bibinfo{journal}{Phys. Rev. B} \textbf{\bibinfo{volume}{52}},
  \bibinfo{pages}{5467} (\bibinfo{year}{1995}).

\bibitem[{\citenamefont{Anisimov et~al.}(1996)\citenamefont{Anisimov, Elfimov,
  Hamada, and Terakura}}]{Anisimov1996}
\bibinfo{author}{\bibfnamefont{V.~I.}~\bibnamefont{Anisimov}},
  \bibinfo{author}{\bibfnamefont{I.~S.}~\bibnamefont{Elfimov}},
  \bibinfo{author}{\bibfnamefont{N.}~\bibnamefont{Hamada}}, \bibnamefont{and}
  \bibinfo{author}{\bibfnamefont{K.}~\bibnamefont{Terakura}},
  \bibinfo{journal}{Phys. Rev. B} \textbf{\bibinfo{volume}{54}},
  \bibinfo{pages}{4387} (\bibinfo{year}{1996}).

\bibitem[{\citenamefont{Wright et~al.}(2001)\citenamefont{Wright, Attfield, and
  Radaelli}}]{Wright2001}
\bibinfo{author}{\bibfnamefont{J.~P.} \bibnamefont{Wright}},
  \bibinfo{author}{\bibfnamefont{J.~P.} \bibnamefont{Attfield}},
  \bibnamefont{and} \bibinfo{author}{\bibfnamefont{P.~G.}
  \bibnamefont{Radaelli}}, \bibinfo{journal}{Phys. Rev. Lett.}
  \textbf{\bibinfo{volume}{87}}, \bibinfo{pages}{266401}
  (\bibinfo{year}{2001}).

\bibitem[{\citenamefont{Cramer and Truhlar}(2009)}]{Cramer2009}
\bibinfo{author}{\bibfnamefont{C.~J.} \bibnamefont{Cramer}} \bibnamefont{and}
  \bibinfo{author}{\bibfnamefont{D.~G.} \bibnamefont{Truhlar}},
  \bibinfo{journal}{Phys. Chem. Chem. Phys.}
  \textbf{\bibinfo{volume}{11}}, \bibinfo{pages}{10757} (\bibinfo{year}{2009}).

\bibitem[{\citenamefont{Reiher}(2007)}]{Reiher2007}
\bibinfo{author}{\bibfnamefont{M.}~\bibnamefont{Reiher}},
  \bibinfo{journal}{Faraday Discuss.} \textbf{\bibinfo{volume}{135}},
  \bibinfo{pages}{97} (\bibinfo{year}{2007}).

\bibitem[{\citenamefont{Vosko et~al.}(1980)\citenamefont{Vosko, Wilk, and
  Nusair}}]{Vosko1980}
\bibinfo{author}{\bibfnamefont{S.~H.} \bibnamefont{Vosko}},
  \bibinfo{author}{\bibfnamefont{L.}~\bibnamefont{Wilk}}, \bibnamefont{and}
  \bibinfo{author}{\bibfnamefont{M.}~\bibnamefont{Nusair}},
  \bibinfo{journal}{Can. J. Phys.} \textbf{\bibinfo{volume}{58}},
  \bibinfo{pages}{1200} (\bibinfo{year}{1980}).

\end{thebibliography}

\appendix
\section{Local DOS}\label{sec:ldos}
\begin{figure}[hb]
\includegraphics[width=8cm]{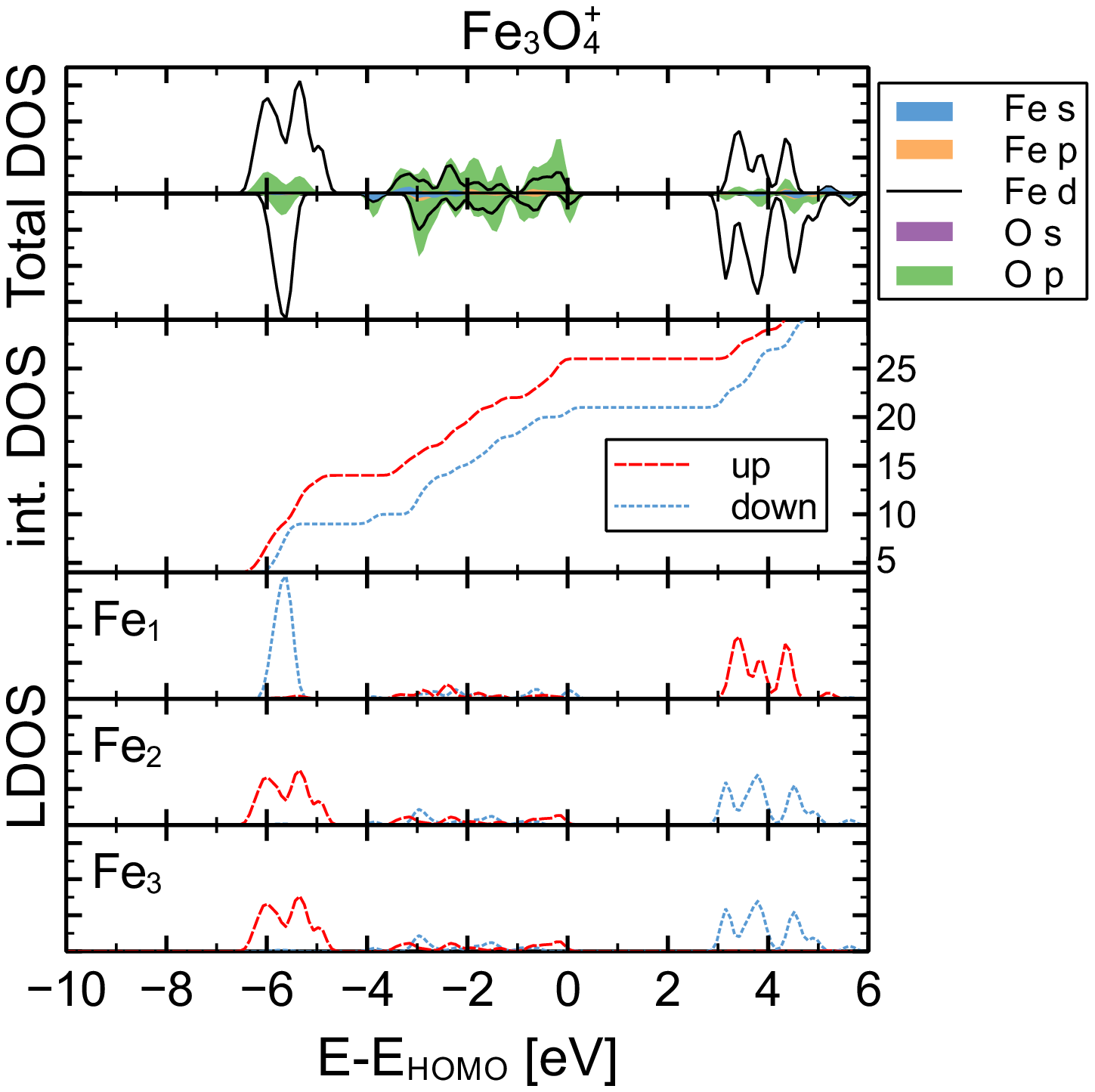}
\caption{(Color online) The total, integrated and local density of states of the \FeOp{3}{4} cluster. 
} 
\label{fig:Fe3O4p_ldos}
\end{figure}

\begin{figure}[h]
\includegraphics[width=8cm]{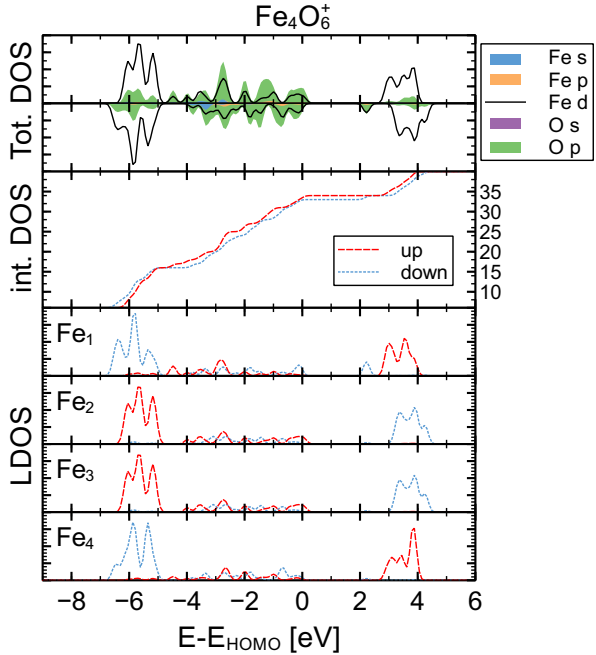}
\caption{(Color online)  The total, integrated, and local density of states of the \FeOp{4}{6} cluster. Fe(1) is tetravalent as is shown in Table~\ref{tab:magnetite_moments}. 
} 
\label{fig:Fe4O6p_ldos}
\end{figure}

\begin{figure}[h]
\includegraphics[width=8cm]{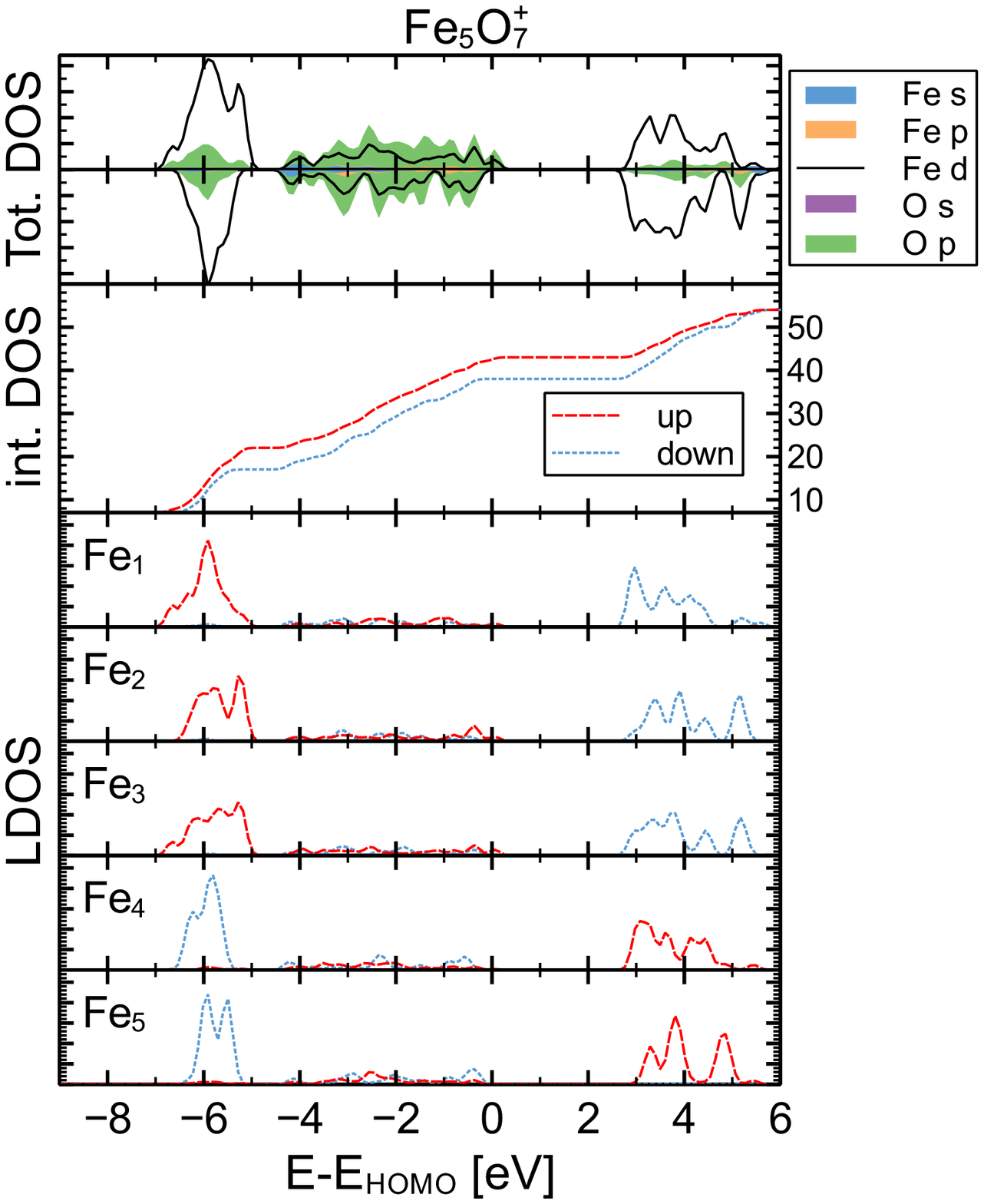}
\caption{(Color online) The total, integrated, and local density of states of the \FeOp{5}{7} cluster. 
} 
\label{fig:Fe5O7p_ldos}
\end{figure}

\begin{figure}[h]
\includegraphics[width=8cm]{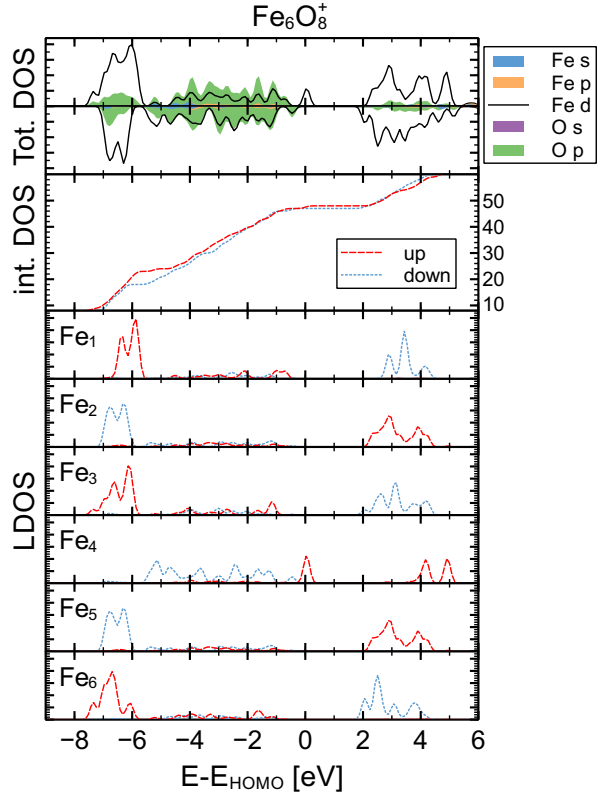}
\caption{(Color online) The total, integrated, and local density of states of the \FeOp{6}{8} cluster. All Fe atoms are trivalent except for Fe(4), which is divalent. } 
\label{fig:Fe6O8p_ldos}
\end{figure}

\begin{figure}[h]
\includegraphics[width=8cm]{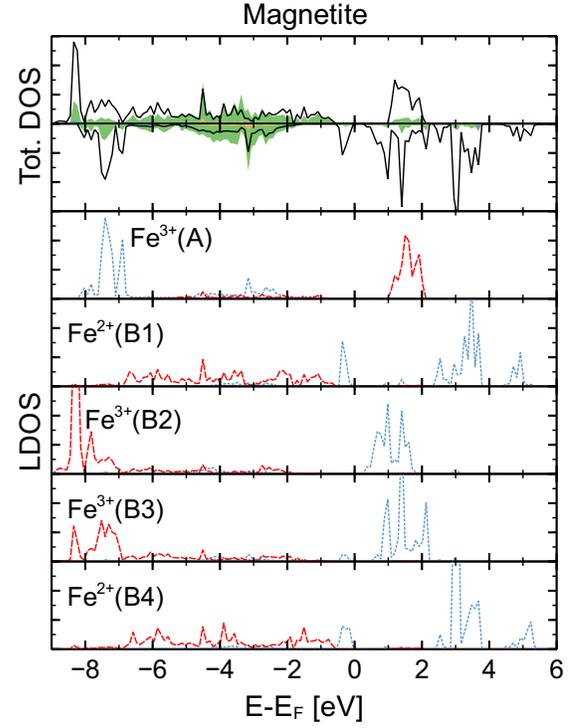}
\caption{(Color online) The total and local density of states of the different Fe atoms in magnetite. The numbering is consistent with Table~\ref{tab:magnetite_moments}. $\text{Fe}^{2+}$ and $\text{Fe}^{3+}$ have a similar LDOS to clusters although the symmetry is very different. } 
\label{fig:Mag_ldos}
\end{figure}

In this appendix we show the integrated and local DOS of the clusters \FeOp{3}{4}, \FeOp{4}{6}, \FeOp{5}{7}, \FeOp{6}{8}, and magnetite. Figures~\ref{fig:Fe3O4p_ldos}, \ref{fig:Fe4O6p_ldos}, \ref{fig:Fe5O7p_ldos}, and \ref{fig:Fe6O8p_ldos} show the total, integrated and local density of states of \FeOp{3}{4}, \FeOp{4}{6}, \FeOp{5}{7}, and \FeOp{6}{8}, respectively. Of these clusters, \FeOp{3}{4} and \FeOp{5}{7} are pure trivalent and the LDOS contains $3d$ peaks at -6~eV and small hybridization between Fe and O. \FeOp{4}{6} contains a single tetravalent Fe atom, with a similar LDOS compared to $\text{Fe}^{3+}$. The ionized electron is not removed from the 3d levels at -6~eV, but from the hybridized levels with oxygen, as can be seen from the integrated density of states. \FeOp{4}{5} and \FeOp{6}{8} contain a single divalent Fe atom, which has a distinct LDOS, in which there are no peaks around -6~eV but strong spin polarized hybridization with oxygen and a single occupied minority level at the HOMO level. Even in bulk magnetite, as is shown in Fig.~\ref{fig:Mag_ldos}, the same features between divalent and trivalent Fe atoms exist. 

\clearpage

\end{document}